%% file: tip_2020_tissier.tex
\documentclass[journal]{IEEEtran}

\usepackage{glossaries,xcolor,graphicx,amssymb,pifont,multirow,diagbox, soul,adjustbox}
\usepackage[colorlinks=true]{hyperref}
\usepackage{booktabs}

\input{acronyms.tex}

\newcommand{\Figure}[1] {Fig.~#1}
\newcommand{\Table}[1] {Table~#1}
\newcommand{\Section}[1] {Section~#1}
\newcommand{\Equation}[1] {Eq.~#1}

\definecolor{mygreen}{rgb}{0.0,0.69,0.31}
\definecolor{myred}{rgb}{0.8745,0.1961,0.1412}

\newcommand{\xmark}{\ding{55}}
\usepackage{cite}

\begin{document}
	
\title{Machine Learning based Efficient QT-MTT Partitioning Scheme for VVC Intra Encoders}

\author{Alexandre~Tissier, Wassim~Hamidouche, Souhaiel~Belhadj Dit Mdalsi,  Jarno~Vanne, Franck~Galpin and~Daniel~Menard
	\thanks{A. Tissier, W. Hamidouche, S. Belhadj Dit Mdalsi and D. Menard are with Univ. Rennes, INSA Rennes, CNRS, IETR - UMR 6164, 20 Avenue des Buttes de Coesmes, 35708 Rennes, France. (emails: whamidou@insa-rennes.fr and dmenard@insa-rennes.fr).}
	\thanks{J. Vanne is with with Tampere University, Korkeakoulunkatu 10, Tampere, 33720, Finland (email: jarno.vanne@tuni.fi).}
	\thanks{F. Galpin is with with InterDigital, Cesson-S\'evign\'e, 35510, France.}
}

\markboth{IEEE Transactions on Circuits and Systems for Video Technology}%
{Shell \MakeLowercase{\textit{et al.}}: Bare Demo of IEEEtran.cls for IEEE Journals}
%

\maketitle

\begin{abstract}
	The next-generation \gls{vvc} standard introduces a new \gls{mtt} block partitioning structure that supports \gls{bt} and \gls{tt} splits in both vertical and horizontal directions. This new approach leads to five possible splits at each block depth. It thereby improves the coding efficiency of \gls{vvc} over that of the preceding \gls{hevc} standard, which only supports \gls{qt} partitioning with a single split per block depth. However, \gls{mtt} also has brought a considerable impact on encoder computational complexity. This paper proposes a two-stage learning-based technique to tackle the complexity overhead of \gls{mtt} in \gls{vvc} intra encoders. In our scheme, the input block is first processed by a \gls{cnn} to predict its spatial features through a vector of probabilities describing the partition at each 4$\times$4 edge. Subsequently, a \gls{dt} model leverages this vector of spatial features to predict the most likely splits at each block. Finally, based on this prediction, only the $N$ most likely splits are processed by the \gls{rd} process of the encoder. In order to train our \gls{cnn} and \gls{dt} models on a wide range of image contents, we also propose a public \gls{vvc} frame partitioning dataset based on existing image dataset encoded with the \gls{vvc} reference software encoder. 
	Our solution relying on the top-3 configuration reaches 47.4\% complexity reduction for a negligible bitrate increase of 0.79\%. A top-2 configuration enables a higher complexity reduction of 70.4\% for 2.49\% bitrate loss. These results emphasize a better trade-off between VTM intra-coding efficiency and complexity reduction compared to the state-of-the-art solutions. The source code of the proposed method and the training dataset are made publicly available at \href{https://github.com/Souhailkudo/VTM_intra_CNN_LGBM_patch}{GitHub}.  	
\end{abstract}

\glsresetall


%
\IEEEpeerreviewmaketitle

\section{Introduction}
\label{sec:intro}
\IEEEPARstart{T}{}he emerging video formats such as 4K/8K and 360-degree videos alongside the explosion of IP video traffic~\cite{cisco_cisco_2019} pushed organizations such as ISO/IEC, ITU-T \gls{jvet} and \gls{aom} to propose new video compression standards. \Gls{aom} developed the AV1 codec released in 2018 as a successor to VP9 and \gls{jvet} developed \gls{vvc} ITU-T H.266 $|$ MPEG-I - Part 3 (ISO/IEC 23090-3)~\cite{9503377, 9689950} in July 2020 as a successor to \gls{hevc}. 

These two standards share the same hybrid video coding structure. Therefore, during standardization, different coding tools were integrated to improve the intra and inter predictions, the in-loop filtering, or to enhance the partitioning of the block of pixels. The selection of specific coding tools leads to different coding efficiencies and computational costs. Several comparison studies were conducted between \gls{vvc} and AV1~\cite{garcia-lucas_rate-distortioncomplexity_2020} based on subjective and objective quality metrics including \gls{psnr} and Video Multi-method Assessment Fusion (VMAF). This latter is a \gls{ml}-based quality metric leveraging detail loss metric, visual information fidelity measure, and averaged temporal frame difference. Both works conclude that \gls{vvc} outperforms AV1 in terms of both objective and subjective quality scores. However, the computational cost of AV1 fluctuates around the \gls{vvc} complexity depending on the considered coding configuration.

The \gls{vvc} reference software, named \gls{vtm}, implements all normative \gls{vvc} coding tools allowing rate-distortion-complexity evaluation and conformance testing. As the successor to \gls{hevc}, the \gls{vtm} implementation brings 25.32\% and 36.95\% bitrate reductions at the expense of a significant increase in encoder computational complexity estimated to 2630\% and 859\% compared to the \gls{hm} 16.22~\cite{bossen_ahg_2020} in \gls{ai} and \gls{ra} configurations, respectively.

Compared to \gls{hevc}, which is based on a \gls{qt} block partitioning, \gls{vvc} integrates a nested \gls{mtt} partitioning scheme allowing in addition to \gls{qt}, horizontal and vertical \gls{bt} and \gls{tt} splits~\cite{huang_block_2021}. This new partitioning scheme is the most efficient tool integrated in \gls{vvc}~\cite{francois_vvc_nodate} with 8.5\% coding efficiency gain reached in \gls{ra} configuration compared to \gls{hevc}. Nevertheless, this coding efficiency improvement is brought at the expense of a significant complexity increase. At each level of the hierarchical partitioning process, up to five splits are tested by the encoder, compared to the one split (i.e., \gls{qt} split) for \gls{hevc}. 
Authors in~\cite{saldanha_complexity_2020} have shown that disabling \gls{bt} and \gls{tt} splits, decreases the encoding time by 91.7\% in \gls{ai}. Therefore, the partitioning process offers the highest opportunity in terms of complexity reduction compared to other coding tools. In~\cite{tissier_complexity_2019}, up to 97.5\% complexity reduction was reported when only the optimal split was tested by the intra encoder at each level of the partitioning process, compared to an exhaustive search. To reach a real-time \gls{vvc} encoding, the complexity of the \gls{qt}-\gls{mtt} partitioning process must be significantly reduced. This work aims to maximize the complexity reduction while minimizing the \gls{bd-br} loss. A large body of literature has investigated the problem of block partitioning for \gls{hevc}~\cite{correa_fast_2015, li_accelerate_2020, mercat_machine_2018, 6862903, 6778776} and VP9~\cite{paul_speeding_2020}. 
For instance, Xu {\it et al.}~\cite{xu_reducing_2018} proposed a \gls{cnn} to predict a hierarchical partition map to avoid exploring improbable block depths.  However, those approaches cannot be directly applied to \gls{vvc} as the new partitioning scheme is significantly different and more complex with the multiplication of available partitions. Therefore,  predicting the optimal block depth becomes more challenging due to the high number of divisions leading to the same depth. 
Recently, a number of complexity reduction methods tailored for \gls{vvc} have emerged through fast encoder strategies~\cite{wieckowski_fast_2019}, probabilistic approach~\cite{wang_probabilistic_2018} or machine learning-based approach~\cite{amestoy_tunable_2020}. These techniques can leverage adaptive resolution as pre-processing~\cite{9690615}, or the prediction of the intra perdition mode~\cite{9328173} and the partitioning mode~\cite{amestoy_tunable_2020}.   
Deep learning techniques~\cite{li_deepqtmt_2020} provided a breakthrough in video coding, especially in the complexity reduction domain. 
Li {\it et al.}~\cite{li_deepqtmt_2020} proposed a \gls{cnn} that predicts the most probable split through a multi-classification at each block depth. One of the drawbacks of this technique is the time overhead related to the \gls{cnn} inference at each depth. Indeed, the  \glspl{cnn} computational complexity must be carefully controlled to not annihilate the gain obtained by the complexity reduction technique. 
Authors in~\cite{tissier_cnn_2020} have recently proposed a \gls{cnn} that processes a block to predict its partition through a vector of probabilities. The average probability over the split edges is then compared to a fixed threshold to decide whether to perform the split or not at each depth. The main drawbacks of this solution are, first, its local decision, which does not leverage all probabilities of the block edges, and second, comparing the average probability of the split edges to a fixed threshold.

In this paper, we propose an efficient complexity reduction technique for the \gls{qt}-\gls{mtt} partitioning. Our approach combines a moderate complexity \gls{cnn} that extracts spatial features of the block of pixels with multi-class classifiers to derive the best partitions for testing in the \gls{rdo} process. A single \gls{cnn} is used to process a $64\times64$ block and predicts the probability of each boundary of all the $4\times4$ pixel blocks within the input $64\times64$ block. At each level of the hierarchical partitioning process, a multi-class classifier is used to predict from the set of boundary probabilities the $N$-most likely splits to explore by the encoder. One classifier is trained for each size of the 16 different sub-blocks. At each depth, the number of explored partitions $N$ can be adjusted from one to the total number of possible splits. Controlling the parameter $N$ allows exploring many trade-offs between complexity reduction and quality loss. The proposed solution with top-$N=3$ configuration reaches 47.4\% complexity reduction for a negligible \gls{bd-br} loss of 0.79\%, while the top-$N=2$ configuration enables in average a complexity reduction of 70.4\% for 2.49\% \gls{bd-br} loss.

The rest of this paper is organized as follows. \Section{\ref{sec:rw}} introduces the partitioning background and the state-of-the-art of complexity reduction techniques. \Section{\ref{sec:pm}} describes the proposed two-stage method that combines two machine learning algorithms. The used dataset to train our models is detailed in~\Section{\ref{sec:dataset}}, followed by the experimental setup and the training process for both \gls{cnn} and \gls{dt} models, presented in~\Section\ref{sec:training_proc}.
\Section{\ref{sec:expe_res}} presents the performance of the two machine learning models and a comparison of our method against state-of-the-art techniques in terms of complexity reduction and \gls{bd-br} loss.  The complexity of the \gls{ml} techniques are assessed and analysed in \Section{\ref{sec:complex_study}}. Finally, \Section{\ref{sec:conclu}} concludes the paper.

\section{Background and Related work}
\label{sec:rw}
In this section, we first describe the frame partitioning in the \gls{vvc} standard, and then we give a brief review of complexity reduction techniques for both \gls{hevc} and \gls{vvc} encoders.   
\subsection{Frame partitioning in \gls{vvc}}
\begin{figure}[]
	\centering
	\includegraphics[width=1\linewidth]{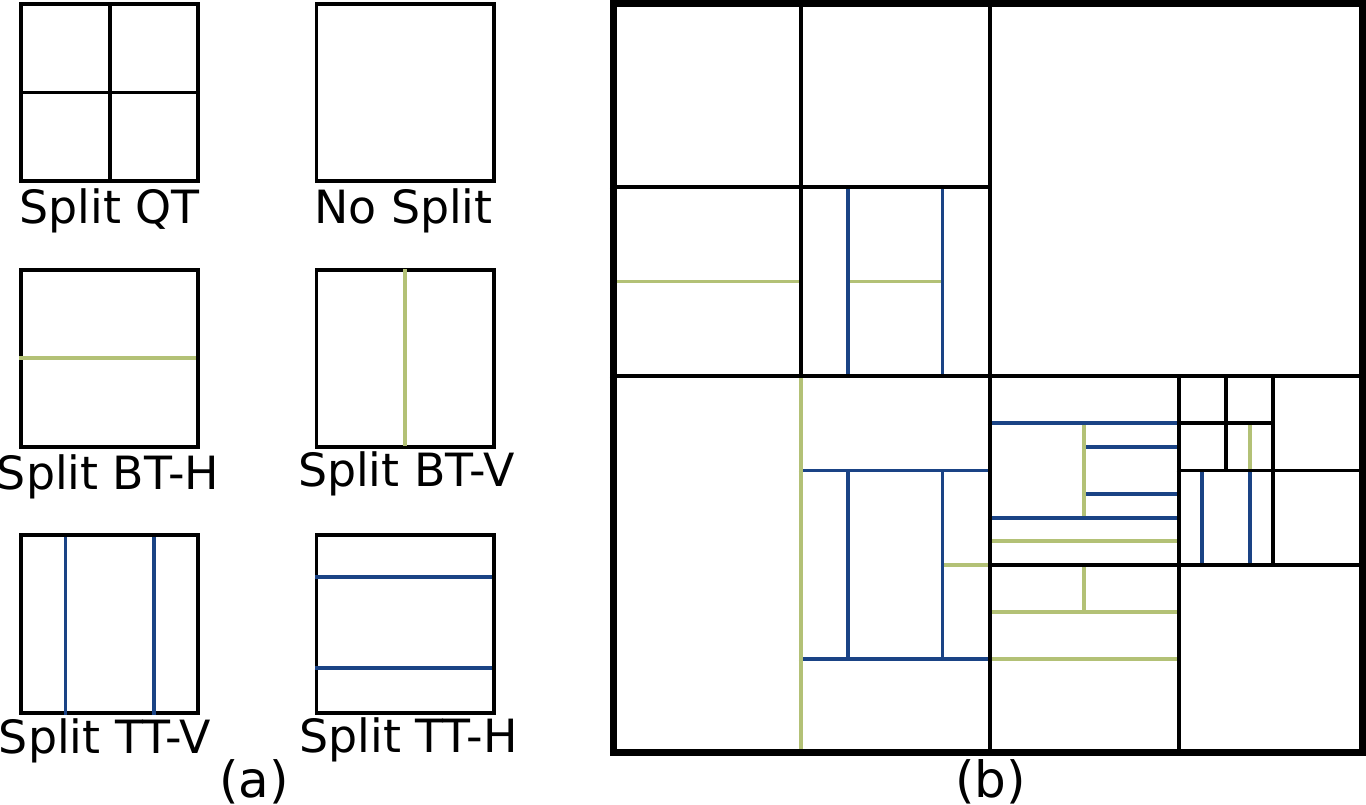}
	\caption{\gls{ctu} partitioning in \gls{vvc}. (a) \gls{vvc} split types. (b) Example of a \gls{ctu} partition in \gls{mtt}.}
	\label{fig:split_example}
\end{figure}

The new block partitioning scheme proposed in \gls{vvc} is the most efficient coding tool integrated into the standard~\cite{wang_extended_2020}. The partitioning process starts from a root block named \gls{ctu}, i.e., a block of $128\times128$ pixels in the \gls{vtm} \gls{ai} configuration. 
The blocks resulting from the partitioning process are named \glspl{cu} and may have a size between $64\times64$ and $4\times4$. \Figure{\ref{fig:split_example}}(a) presents the split modes supported by \gls{vvc}.  As \gls{hevc}, \gls{qt} divides a \gls{cu} into four equal sub-\glspl{cu}.  Moreover, \gls{vvc} allows a rectangular shape for \gls{cu} with its novel splits \gls{bt} and \gls{tt}. The \gls{bt} divides a \gls{cu} into two sub-\glspl{cu} while the \gls{tt} divides a \gls{cu} into three sub-\glspl{cu} with the ratio 1:2:1.  Both \gls{bt} and \gls{tt} can split a \gls{cu} horizontally or vertically. \Figure{\ref{fig:split_example}}(b) presents the splits of a \gls{ctu} processed by the \gls{vtm} \gls{rdo}.  The \gls{rdo} process relies on an exhaustive search that calculates the \gls{rd}-cost for each \gls{cu}, then selects the \gls{ctu} partition that results in the lowest \gls{rd}-cost. In \gls{ai} configuration, the \gls{vtm} forces the first split of the \gls{ctu} to be a \gls{qt}. Additional constraints are also applied to \glspl{cu}, such as  \gls{qt}  split is not allowed after a \gls{bt} or a \gls{tt} split. Excluding the \gls{vtm} restrictions, the encoder performs all possible splits on each \gls{cu} to select the optimal partition with the lowest \gls{rd}-cost. 

\subsection{Existing techniques for encoder complexity reduction}

A brief review of complexity reduction methods proposed to speed up the reference software encoders of the \gls{hevc} and \gls{vvc} standards is provided in this section.
Authors in~\cite{tissier_complexity_2019} have studied the complexity reduction opportunities in the \gls{vtm}3.0 under the \gls{ai} configuration.  
Several tools have been identified to reduce the encoding complexity, such as the partitioning process, the intra mode prediction, and multiple transform selection. 
This study showed that the partitioning process has the highest impact on encoder computational complexity with 97.5\% complexity reduction, followed by the intra mode prediction with 65.2\% and the multiple transform selection with 55.2\%. 
This study motivated the work of this paper that tackles the complexity reduction of the partitioning process. Meanwhile, several works have studied the complexity reduction of the intra mode prediction~\cite{zhang_fast_2020} and the multiple transform selection~\cite{fu_two-stage_2019}. To reduce the complexity of the partitioning process, state-of-the-art solutions rely on different approaches to compute features relevant to this problem, including handcrafted and learning-based techniques. The computed features are then fed to a classifier such as \gls{dt}, \gls{svm}, or dense layers. 
It should be noted that the performance of the following techniques is provided under the \gls{ai} coding configuration unless otherwise specified.

\subsubsection{\gls{hevc} complexity reduction techniques}
The complexity reduction of the \gls{hevc} recursive block partitioning has been widely investigated in the literature. 
Min {\it et al.}~\cite{biao_min_fast_2015} defined a complexity score metric that predicts the spacial complexity of a block.
The complexity score is computed as the l1 norm of the differences between the luminance pixel value at a position and the mean luminance value.
This complexity score is computed for the two sub-blocks obtained by dividing the block into horizontal, vertical, or two diagonals.
Then, the  difference between the two complexity values of the resulting sub-blocks is computed. 
This value is then compared to a predefined threshold to decide whether the block should be split, not split, or undetermined.
This solution reaches 52\% of complexity reduction at the expense of a slight \gls{bd-br} increase of 0.8\%. Correa {\it et al.}~\cite{correa_fast_2015} used three sets of a decision tree trained with features such as \gls{rd}-cost, gradient, the sum of pixels, or variance of pixels. These decision trees predict whether the \gls{cu}, \gls{pu}, or \gls{tu} must be split or not. By predicting these partitioning specific to \gls{hevc}, this solution achieves a 65\% of computational complexity reduction for 1.36\% \gls{bd-br} loss.
\glspl{cnn} have already been exploited for \gls{hevc} encoding complexity reduction~\cite{xu_reducing_2018, li_accelerate_2020}.
Xu {\it et al.}~\cite{xu_reducing_2018} first used a \gls{cnn} to predict a hierarchical \gls{cu} partition map which provides an efficient representation of the \gls{cu} partitioning in intra mode. 
A \gls{lstm} network was then integrated to predict the partition in inter prediction mode.
In \gls{ai} configuration, this solution reduces the complexity by 62\% for 2.25\% \gls{bd-br} increase. 
Under the low delay P coding configuration, it reaches 54.2\% of computational complexity reduction for 1.5\% \gls{bd-br} increase. 
Li {\it et al.}~\cite{li_accelerate_2020} claimed a real-time \gls{ctu} partition prediction based on their previous proposed complexity reduction method presented in~\cite{xu_reducing_2018}.
They simplified the \gls{cnn} by pruning the weights at different levels to perform multiple approximations. 
These different configurations allow complexity control at \gls{ctu} level by selecting the proper \gls{cnn}. An estimation of the \gls{cnn} run time is performed at the frame level to enhance the stability of complexity control. The \gls{cnn} run time speed up is improved by 17 to 20 times with a complexity control error of 2\%.

\subsubsection{\gls{jem} complexity reduction techniques}
The \gls{jem} software was developed in early 2014 to study the potential coding gain behind developing a new standard with coding performance beyond \gls{hevc}. 
The \gls{jem} software is based on \gls{hm} with new coding tools such as \gls{mts}, and \gls{bt} proposed to enhance the partitioning efficiency at the cost of higher computational complexity.
Wang {\it et al.}~\cite{wang_probabilistic_2018} proposed a novel \gls{rd}-cost estimation scheme relying on the motion divergence field. Based on the estimated \gls{rd}-cost, a probabilistic framework is developed to skip unnecessary splits.
The proposed algorithm reduces the complexity by 54.7\% for a 1.15\% \gls{bd-br} increase in \gls{ra} configuration. 
The same authors proposed in~\cite{wang_effective_2017} the choice of a dynamic parameter at \gls{ctu} level based on neighboring partitions as the first step. In the second step, \gls{qt} and \gls{bt} decision tree classifiers predict the probability of the different splits to derive the proper partition. 
These two techniques enable 67.6\% complexity reduction for a 1.34\% \gls{bd-br} increase in \gls{ai} configuration.
Jin {\it et al.}~\cite{jin_cnn_2017} designed a multi-classification \gls{cnn} that predicts the partition depth of a $32\times32$ \gls{cu}. 
This method enables skipping unnecessary splits for the partition depth predicted outside the candidate depth range. As a result, the encoder computational complexity is reduced by 42.8\% for a \gls{bd-br} increase of 0.65\%.

\begin{table*}[]
	\centering
	\caption{Main features of the existing state-of-the-art complexity reduction techniques in All Intra coding configuration. CR: Complexity reduction}
	\label{tab:rw}
	\begin{tabular}{l|c|c|c|c|c|c|c}
	\toprule
		\multicolumn{1}{c|}{Solution} & Handcrafted & Decision Tree & Neural network & Software & CR (\%)$\uparrow$ & BD-BR (\%)$\downarrow$ & CR/BD-BR ratio$\uparrow$ \\ 	\midrule
		Biao {\it et al.}~\cite{biao_min_fast_2015}                           & \checkmark           & \xmark                & \xmark              & HM 10.0      & 52.0\%                        & 0.80\%    &   65.00  \\ 
		Xu {\it et al.}~\cite{xu_reducing_2018}                             & \xmark           & \xmark                & \checkmark              & HM 16.5      & 62.0\%                        & 2.25\%       & 27.55  \\ 
		Wang {\it et al.}~\cite{wang_effective_2017}                           & \checkmark           & \checkmark                & \xmark              & HM 13.0 QTBT & 67.6\%                      & 1.34\%   & 50.44    \\ 
	\midrule
		Jin {\it et al.}~\cite{jin_cnn_2017}                            & \xmark           & \xmark                & \checkmark              & JEM 3.1      & 42.8\%                      & 0.65\%       &  65.84 \\
			\midrule
		Yang {\it et al.}~\cite{yang_low_2019}                           & \checkmark           & \checkmark                & \xmark              & VTM 2.0      & 52.0\%                     & 1.59\%    &  32.70 \\ 
		Chen {\it et al.}~\cite{chen_fast_2020}                           & \checkmark           & \checkmark                & \xmark              & VTM 4.0      & 51.2\%                     & 1.62\%  &  31.60   \\ 
		Li {\it et al.}~\cite{li_deepqtmt_2020}                             & \xmark           & \xmark                & \checkmark              & VTM 7.0      & 45.8\%                     & 1.32\%   & 34.69   \\ 
		Zhao {\it et al.}~\cite{zhao_adaptive_2020}                           & \checkmark           & \xmark                & \checkmark              & VTM 7.0      & 39.4\%                     & 0.87\%    & 45.28  \\ 
		Saldanha {\it et al.}~\cite{saldanha_configurable_2021}                             & \checkmark           & \checkmark                & \xmark              & VTM 10.0      & 48.8\%                        & 1.01\%      & 48.31  \\ 
	Tissier {\it et al.}~\cite{tissier_cnn_2020} & \xmark  & \xmark & \checkmark &  VTM 10.2 & 54.0\% & 1.40\% & 38.57 	\\ 
	Ours (Top-3) &  \xmark &  \checkmark & \checkmark  & VTM 10.2 &	47.4\% & 0.79\% &  60.00 \\ \bottomrule
	\end{tabular}
	\vspace{-4mm}
\end{table*}

\subsubsection{\gls{vvc} complexity reduction techniques}
Although  \gls{vvc} was recently standardized, several techniques have already been proposed to tackle the problem of encoder complexity. 
Indeed, the extension of the partitioning process with \gls{qt}, \gls{bt}, and \gls{tt} splits considerably increases the computational encoding complexity.
Predicting a split at each \gls{cu} becomes a real challenge due to the number of partition possibilities that have increased significantly compared to \gls{hevc}.
This situation raises the need for a lightweight partitioning process that decreases the encoder complexity while preserving its coding efficiency for live applications. 
Lei {\it et al.}~\cite{lei_look-ahead_2019} proposed a two-step look-ahead prediction method for the intra prediction mode and the partitioning process.
This solution computes the rate-distortion cost of only 7 intra modes out of 67, and if a \gls{cu} has multiple partitions, the \gls{rd}-cost of partitioned \gls{cu} is computed from the parent \gls{cu}.
Moreover, this solution approximates the \gls{rd}-costs of different partition directions in order to skip unnecessary directions. Intra mode technique combined with partitioning prediction technique allows a computational complexity reduction of 45.8\% for a 1.03\% \gls{bd-br} increase.
Amestoy {\it et al.}~\cite{amestoy_tunable_2020} proposed a cascade framework through random forest classifiers to determine the probability of each split. 
To improve the accuracy of their classifiers, the impact of each feature is evaluated, such as \gls{qp}, variance, and mean of the block gradients.
Furthermore, the thresholds applied to the different classifiers are optimized.
A risk interval was proposed to limit the \gls{rd}-cost increase by computing both splits of the classifier output when the probability falls in this risk interval.
In \gls{ra} configuration, this solution enables from 30.1\% to 61.5\% of complexity reduction for respectively 0.61\% to 2.22\% \gls{bd-br} increase depending on the risk interval configuration.
Another cascade framework was proposed by Yang {\it et al.}~\cite{yang_low_2019} with one binary classifier for each split mode. The authors used three categories of features, including global texture information,  local texture information, and context information, as input for their classifiers.
A fast intra mode decision using a one-dimensional gradient descent search is combined with the \gls{ctu} structure decision.
This fast partitioning solution achieves 52\% of complexity reduction for a 1.59\% \gls{bd-br} increase. The intra mode decision technique enhances the complexity reduction to 62.5\% for 1.93\% \gls{bd-br} loss.
Chen {\it et al.}~\cite{chen_fast_2020} also used a supervised learning method. 
Instead of a cascade framework, they designed six \gls{svm} models depending on the \gls{cu} sizes, which are trained to skip vertical and horizontal splits.
\gls{svm} features are derived from entropy, texture contrast, and Haar wavelet of the current \gls{cu}.
This solution reaches a 51.2\% of computational complexity reduction for a 1.62\% \gls{bd-br} increase.
Li {\it et al.}~\cite{li_deepqtmt_2020} proposed a deep learning approach to predict the \gls{ctu} partition with a binary or multi-classification at each \gls{cu} depth.
To train the \gls{cnn}, they designed an adaptive loss function that defines penalty weights based on a split proportion that penalizes the high difference between the \gls{rd}-cost for the split predicted and the minimum \gls{rd}-cost of the parent \gls{cu}.
Based on the prediction accuracy of the \gls{cu} size, an adaptive threshold is compared to the output of the \gls{cnn}.
This technique enables the reduction of the complexity by 45.8\% for a 1.32\% \gls{bd-br} increase.
Another \gls{cnn} model is proposed by Zhao {\it et al.} in~\cite{zhao_adaptive_2020}. 
First, the standard deviation of \gls{cu} pixels is compared to an adaptive threshold based on \gls{qp} and \gls{cu} depth in order to classify a block into complex or homogeneous \gls{cu}.
The \glspl{cu} defined as homogeneous are no more split. 
Second, for complex \gls{cu}, a \gls{cnn} is trained to predict whether or not the current \gls{cu} must be early terminated.
This solution achieves a 39.4\% computational complexity reduction for a 0.87\% \gls{bd-br} increase.
Saldanha {\it et al.}~\cite{saldanha_configurable_2021} presented a configurable partitioning decision using a \gls{lgbm} model.
Five \gls{lgbm} binary classifiers are trained offline, exploiting handcrafted features such as \gls{qp}, width, or variance.
The classifiers predict a split probability which is compared to a threshold  to skip the split.
This technique obtains several trade-offs with a complexity reduction from 35.2\% to 61.3\% and a \gls{bd-br} increase from 0.46\% to 2.43\%.

\Table{\ref{tab:rw}} summarizes the features and performance of the \gls{vvc} complexity reduction techniques mentioned above.  Compared to the state-of-the-art solutions, the contributions of our paper are summarized as follows: 1) Training decision tree classifiers by \gls{cu} size instead of performing a simple thresholding decision.   2) Dataset balancing and enhancing with integrating more screen content video sequences.  3) Assess the complexity of the \gls{cnn} and \gls{dt} models on different CPU and GPU platforms as a first step toward integrating our model in professional \gls{vvc} encoders. The proposed solution outperforms state-of-the-art methods enabling the highest ratio  between complexity reduction and \gls{bd-br} loss.

\section{Proposed two-stage \gls{vvc} partition prediction solution}
\label{sec:pm}
\begin{figure*}[]
	\centering
	\includegraphics[width=1\linewidth]{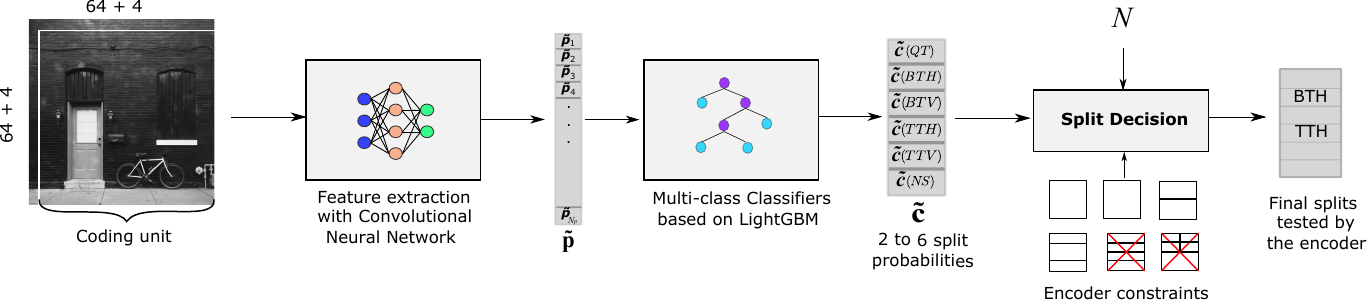}
	\caption{Workflow diagram of the proposed block partitioning scheme for gls{vvc}  in \gls{ai} coding configuration.  A CNN first processes the input luminance block to predict $\mathbf{\tilde{p}}$, a vector of $N_{p}$ probabilities describing all edges at each $4\times4$ sub-block. The vector $\mathbf{\tilde{p}}$ is then processed by a decision tree LightGBM to predict the probabilities of the six partitioning modes through the vector $\mathbf{\tilde{c}}$. The top-N splits with highest probabilities are tested by the \gls{rd} process of the encoder to select the optimal split in terms of \gls{rd}-cost.}
	\label{fig:global_scheme}
\end{figure*}

As described in \Section{\ref{sec:intro}}, \gls{vvc} introduces \gls{bt} and \gls{tt} splits at the cost of a high increase in computational complexity. 
In addition, this recursive partitioning process computes the rate-distortion cost for a set of coding tools for each \gls{cu}. 
The proposed partitioning prediction technique is composed of a \gls{cnn} for spatial features extraction followed by a set of multi-class classifiers based on \glspl{dt} to predict the appropriate partitioning decision at different depths of the partitioning tree. 
\Figure{\ref{fig:global_scheme}} illustrates the flow diagram of the proposed  partitioning prediction solution.

\subsection{Overall presentation}

\begin{figure}[]
	\centering
	\includegraphics[width=0.7\linewidth]{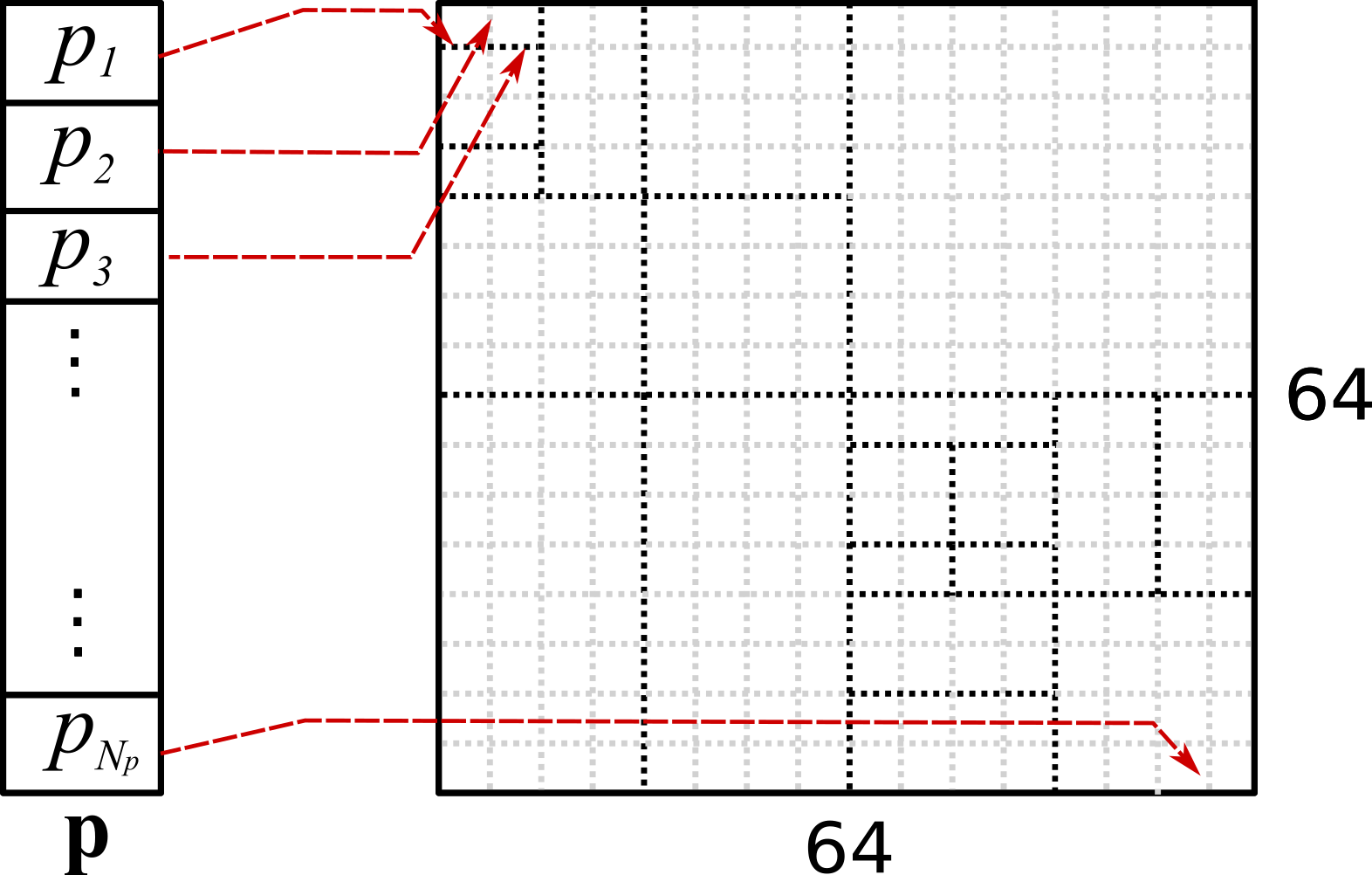}
	\caption{Correspondence between the \gls{cnn} output vector and a $64\times64$ \gls{cu} partition.}
	\label{fig:cnn_output}
\end{figure}

\begin{figure*}[]
	\centering
	\includegraphics[width=0.8\linewidth]{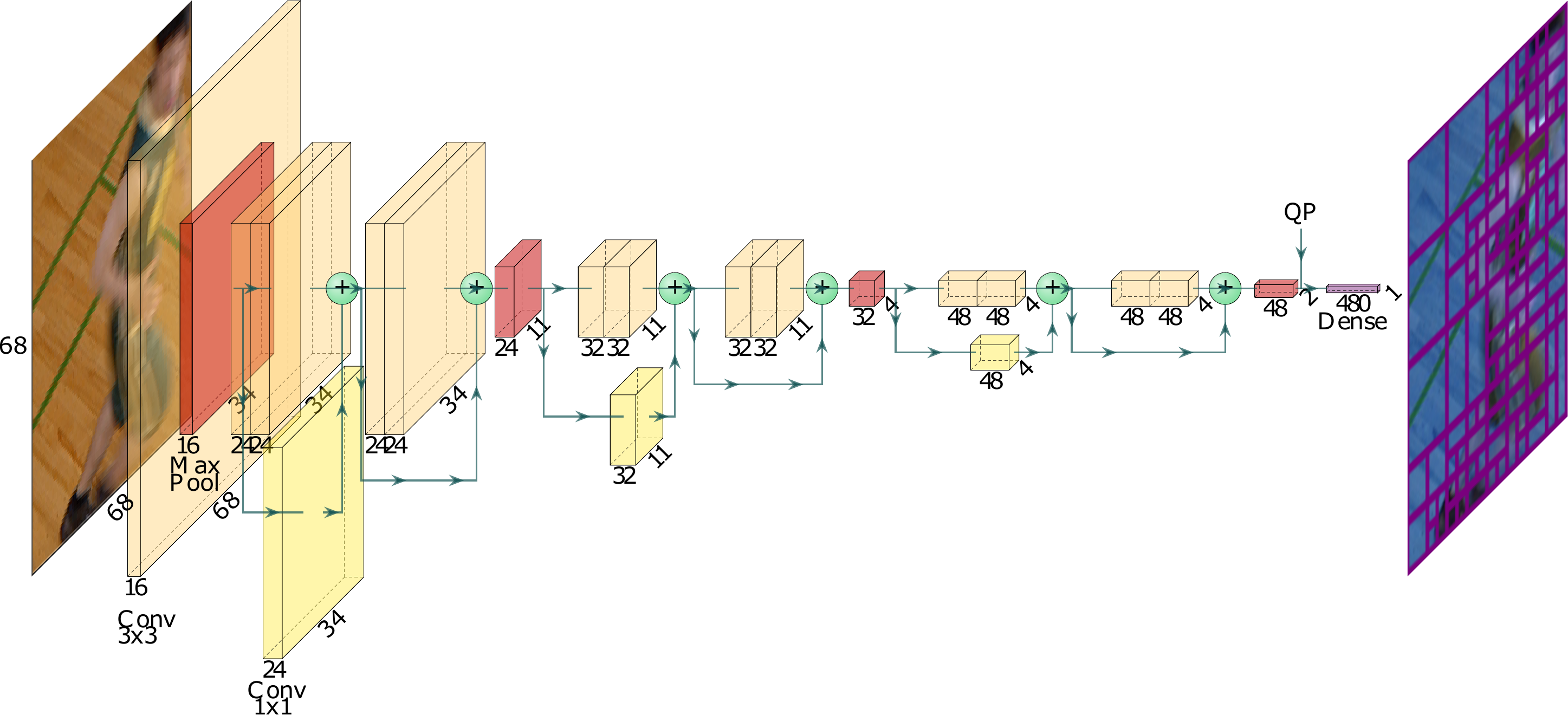}
	\caption{The \gls{cnn} architecture with convolution layers in orange and yellow, max-pooling layers in red and fully connected layer in purple.}
	\label{fig:cnn_explanation}
\end{figure*}
Our method is a top-down solution that early terminates unlikely splits, i.e., it follows the hierarchical process of the \gls{vvc} encoder and skips split modes that have a low probability of belonging to the optimal partitioning. 
Our complexity reduction method comprises two components: spatial features extraction and \gls{dt} classifiers. 
The features extraction block consists of a \gls{cnn} that processes an input luminance block $\mathbf{B}$ to predict a vector of probabilities $\mathbf{\tilde{p}}$ of splits at the $4\times4$ block edges: 
\begin{equation}
\mathbf{\tilde{p}} = f_\theta (\mathbf{B}). 
\end{equation}
where $f_\theta$ is a parametric function of the \gls{cnn} with trainable parameters $\theta$ and $\mathbf{B}$ is the input block of size $68\times68$. 

The block $\mathbf{B}$ consists of the current \gls{cu} of size $64\times64$ padded with four rows and four columns on the top and left of the block resulting in a block size of $68\times68$. 
The left and top neighbor pixels are used as reference samples in the intra prediction through the \gls{mrl} intra prediction~\cite{bross_multiple_2018}. Indeed, their pixels are used to derive the intra mode of the current block. Therefore, it is necessary to include these samples in the features extraction stage, i.e., the \gls{cnn}. 
The \gls{cnn} predicts for each $N_{B}\times N_{B}$ \gls{cu} a vector of $N_{p}$ probabilities of a split at each $4\times4$ edge of the block. The length $N_{p}$ of the predicted probabilities vector $\mathbf{\tilde{p}}$ is computed as follows:
\begin{equation}
N_{p} = \frac{N_{B}}{2} \left( \frac{N_{B}}{4}-1 \right).
\end{equation}

In the case of \gls{cu} size of $64\times64$, $N_{p}$ is the length of the vector $\mathbf{\tilde{p}}$ which is equal to 480. \Figure{\ref{fig:cnn_output}} presents the connections between the probability vector components and the 480 $4\times4$ edges of an input \gls{cu}. 
As this figure highlights, the first $\mathbf{p}_{1}$ and the third $\mathbf{p}_{3}$ components of the vector correspond to the horizontal bottom edges of the first and second $4\times4$ blocks, respectively. For an accurate partitioning prediction, these two components must have a value (probability) close to 1 as a \gls{tt} split is performed to split the sub-block. The last component of the vector $\mathbf{p}_{N_p}$ has a value (probability) close to 0 since no split is performed at this final @vertical edge of the \gls{cu} as illustrated in \Figure{\ref{fig:cnn_output}}.

The classifiers are then fed with the probabilities vector $\mathbf{\tilde{p}}$ derived from the \gls{cnn} to predict the split decision to perform at each tree depth     
\begin{equation}
\mathbf{\tilde{c}} = g_{\omega_i} (\mathbf{\tilde{p}}),  \; \;  \forall i \in \{1, \dots, M \},
\end{equation}
where $g_{\omega_i}$ is a parametric function of classifier $i$,  ${\omega_i}$ is its trainable parameters, and $M$ is the number of considered classifiers. 

A separate multi-class classifier based on \gls{dt} model is applied for each \gls{cu} size. 
The classifier takes as an input the probability vector $\mathbf{\tilde{p}}$ and predicts a vector $\mathbf{\tilde{c}}$ of six probabilities that correspond to the six possible split modes performed by the \gls{vvc} encoder at each tree depth. 
This approach results in $M$ separate \gls{dt} models that are trained separately to enhance the prediction performance and enable better model convergence the  with a reduced number of trainable parameters. 
Once the split probabilities are derived for the \gls{cu}, a selection of the highest probabilities is made based on the selected configuration. The configuration specifies the $N$ value, which defines the number of tested splits by the encoder. Thus, the encoder tests only the $N$ first splits corresponding to the highest probabilities (Top-N) predicted by the \gls{dt} model.       
The encoder then skips the remaining splits with lower probabilities than the top-N candidates to reduce the encoding computational complexity. 
The proposed spatial features extraction \gls{cnn} model and the \gls{dt} classifiers are described in more detail in the following two sections. 

\subsection{Spatial features extraction \gls{cnn} model}
\Figure{\ref{fig:cnn_explanation}} presents the adopted \gls{cnn} architecture, which is inspired by the ResNet network~\cite{he_deep_2015}. 
The orange layers represent convolution layers with $3\times3$ kernel (Conv $3\times3$), whereas the yellow layers denote convolutions with $1\times1$ kernel (Conv $1\times1$) that transform the input feature map matrix to match the dimension of the next layer which are then summed up (green plus). 
The red layers correspond to the max pooling (Max Pool) that subsample the input features map with a window of $2\times2$ by selecting the maximum over four values. 
The last layer in purple is a fully connected layer (Dense) that predicts the 480 components vector. 
The sigmoid activation function is used to predict the output values within the interval $[0, 1]$. 
All these layers result in a network with 226,088 trainable parameters. 
The input consists of  $68\times68$ luminance pixels of the \gls{cu} currently processed plus the \gls{qp} value that highly influences the final partition. 
The \gls{qp} is provided as an external input to the last fully connected layer.  
The \gls{qp} parameter is crucial for saving the memory bandwidth as it leads to only one model shared for all \gls{qp} values instead of adopting one model by \gls{qp} value.  
Therefore, our model can be efficiently used with a rate control mechanism that adapts the \gls{qp} value to the target bandwidth. 
The output is a 480 components vector representing the fine-grain partitions ($4\times4$) of the $64\times64$ \gls{cu}. 
This solution has the advantage of predicting the whole \gls{cu} spatial features in one shot, which is very convenient for reducing the complexity overhead and latency introduced by this step. 
Moreover, the architecture of the network is less deep than state of the art \gls{cnn} architectures~\cite{tan_efficientnet_2019, simonyan_very_2015} and thus will require less time to predict the output vector.

\subsection{Multi-class classifiers based on \gls{dt} models}
The decision approach adopted in our previous work~\cite{tissier_cnn_2020} relied only on the probability at the spatial location of a specific split.  The pixel level characteristics of the input \gls{cu} are well exploited by the \gls{cnn} and are represented in the output vector $\mathbf{\tilde{p}}$. Nevertheless,  the partitioning decision considered in \cite{tissier_cnn_2020} uses only information in the split edges (local), while the \gls{dt} may benefit from information of all edges in the \gls{cu} (the whole vector of probabilities, i.e., global). The maximum convolution kernel size may limit the \gls{cnn} extracting these global features in a \gls{cu}.

The vector of probabilities $\mathbf{p}$ can be considered as spatial features relevant to the block partitioning process.  Therefore, the \gls{cnn} inference is carried out once for each \gls{cu} of size $64\times64$ to predict the corresponding probability vector $\mathbf{p}$, then a specific \gls{dt} model processes this vector at each level of the partitioning tree to derive a set of $N$ more likely splits to explore by the encoder.     
To predict split probabilities at various \gls{cu} depths, we consider multiple models covering all possible sizes of the rectangular sub-blocks from $64\times64$ to $4\times4$ excluded.  Table~\ref{tab:size_table} illustrates that the sixteen \gls{cu} sizes can be further split into two to six different partition modes, including the no split mode. 
The \gls{dt} model is fed with the probability vector $\mathbf{\tilde{p}}$ and the \gls{qp} value. 
The probability vector is then cropped into a sub-vector that includes only the probabilities of the sub-block edges. 
The \gls{dt} model performs a multi-class classification by predicting a probability vector $\mathbf{\tilde{c}}$ of six components corresponding to the probabilities of the six possible splits. Therefore, the probabilities of non-possible splits are set to zero during the training process. 

Several machine learning models were tested to solve this multi-class classification problem including \gls{dt}, random forest, \gls{svm} with different kernels, and \gls{lgbm} model~\cite{ke_lightgbm_2017}. However, this latter was considered based on its excellent classification performance and low complexity at both training and inference stages.

\begin{table}[!ht]
	\centering
	\caption{Possible splits according to the \gls{cu} size}
	\label{tab:size_table}
	\begin{tabular}{|c|c|c|c|c|c|}
		\hline
		\backslashbox{Height}{Width} & 64 & 32 & 16 & 8 & 4 \\ \hline
		64 & QT & \multicolumn{4}{c|}{-} \\ \hline
		32 & \multirow{4}{*}{-} & All & \gls{bt}, \gls{tt} & \multirow{2}{*}{\begin{tabular}[c]{@{}c@{}}\gls{bt}\\ \gls{tt}H\end{tabular}} & \multirow{2}{*}{\begin{tabular}[c]{@{}c@{}}\gls{bt}H\\ \gls{tt}H\end{tabular}} \\ \cline{1-1} \cline{3-4}
		16 &  & \gls{bt}, \gls{tt} & All &  &  \\ \cline{1-1} \cline{3-6} 
		8 &  & \multicolumn{2}{c|}{\gls{bt}, \gls{tt}V} & \gls{bt} & \gls{bt}H \\ \cline{1-1} \cline{3-6} 
		4 &  & \multicolumn{2}{c|}{\gls{bt}V, \gls{tt}V} & \gls{bt}V & - \\ \hline
	\end{tabular}
\end{table}


\section{Proposed dataset for training}
\label{sec:dataset}
In this section, we present the dataset in its soft and hard representations and its balancing process.

\subsection{Dataset for the learning process}
\label{subsec:dataset}

\begin{table}[]
	\centering
	\caption{Breakdown of our dataset by resolution.}
	\label{tab:n_images}
	\begin{tabular}{l|c|c|c|c|c|c|c} 
		\toprule
		Resolution      & 240p & 480p & 720p & 1080p & 4K  & 8K & Total \\ 	\midrule
		Nb images & 500  & 500  & 579  & 2557  & 654 & 418 & 5208 \\ 	\bottomrule
	\end{tabular}
\end{table}

The lack of a public dataset providing encoded blocks with the \gls{vtm}, and their corresponding partitions drives us to construct our training dataset to optimize the proposed models'  weights. 
Our work focuses on \gls{ai} configuration, thus the temporal relationship between adjacent frames is not considered.
Five public image datasets were selected including Div2k~\cite{agustsson_ntire_2017}, 4K images~\cite{makov_dataset_2019}, jpeg-ai~\cite{jtc_call_2020}, HDR Google~\cite{hasinoff_burst_2016} and flickr2k~\cite{timofte_ntire_nodate}. 
The resulting dataset presents a high diversity of still image contents. 
However, since these datasets include more images in high-resolution (Full HD and 4K resolutions), a set of high-resolution images are downscaled to lower resolutions with a bilinear filter and added to the dataset.  The resulting dataset includes around 5208 images at different resolutions as detailed in~\Table{\ref{tab:n_images}}, which gives the number of pictures per resolution.
Images of similar resolution are then concatenated to build a pseudo-video sequence. This latter is encoded with the \gls{vtm} encoder in \gls{ai} configuration at different \gls{qp} values, \glspl{qp} $\in \{ 22, \,  27, \,  32,  \, 37 \}$. 

It should be noted that the \gls{vtm} encoder includes multiple speed-up techniques, reducing the complexity brought by the \gls{vvc} partitioning process~\cite{wieckowski_fast_2019}. 
However, to achieve a high coding efficiency by testing more partitioning configurations, these speed-up techniques have been disabled to build our dataset.
Compared to the \gls{vtm} anchor, disabling these speed-up techniques 
leads to more accurate partitioning configurations, enhancing the coding efficiency.
Nevertheless, a higher encoding time is needed to create the ground truth, but only one encoding pass is required, so increasing the encoding time is not critical at this stage. 
The \gls{vtm} in \gls{ai} configuration relies on the dual-tree tool that performs separate partitioning for luminance and chrominance components.
The partitioning information of both components is recorded, while only the prediction of luminance partitioning is considered in this paper since it takes most of the encoding complexity with more than 85\% of the total encoding time~\cite{saldanha_complexity_2020}.

\begin{figure}[]
	\centering
	\includegraphics[width=1\linewidth]{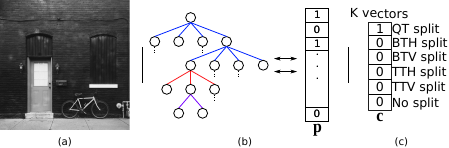}
	\caption{Representation of our dataset. (a) is a luminance $68\times68$ input block $\mathbf{B}$. (b) is the optimal partitioning of the block represented by a tree and transformed to the soft representation, i.e. a 480 elements vector $\mathbf{p}$. (c) is the hard representation, $K$ vectors $\mathbf{c}$ of 6 probabilities, that define the optimal split for each of the $K$ blocks in the tree. $K$ is the decision tree depth.}
	\label{fig:cnn_scheme_input_output}
\end{figure}

The optimal partitions computed by the \gls{vtm} encoder is first saved as a tree. Therefore, to integrate the dataset into our proposed method, two data representations are defined as soft and hard representations, as illustrated in \Figure{\ref{fig:cnn_scheme_input_output}}.
\Figure{\ref{fig:cnn_scheme_input_output}}-(a) shows a block $\mathbf{B}$ processed by the \gls{vtm} encoder which is a $64\times64$ luminance block plus 4 rows and columns on top and left of the block.
These extra pixels are required for the \gls{mrl} intra prediction tool.
\Figure{\ref{fig:cnn_scheme_input_output}}-(b) illustrates the partition tree of the this block which is converted to a 480 components vector.
This soft representation of the tree depicts the $64\times64$ block luminance partition in $4\times4$ blocks in a single vector $\mathbf{p}$. 

The hard representation of our dataset is a succession of $K$ vectors that defines all splits at each depth.
\Figure{\ref{fig:cnn_scheme_input_output}}-(c) presents one of those vector $\mathbf{c}$ which defines the optimal split for a specific \gls{cu} size.
The vector size is set to six as the maximum number of splits defined by \gls{vvc}.
For instance, the $64\times64$ \gls{cu} size has only two possible splits with \gls{qt} and no split, and thus the four remaining components of the vector are set to zero. Instead, for a \gls{cu} size of $32\times32$, all splits are available, so the vector size is 6 with \gls{qt}, the two \glspl{bt}, the two \glspl{tt} and no split.

\subsection{Dataset processing}
\begin{figure}[]
	\centering
	\includegraphics[width=1\linewidth]{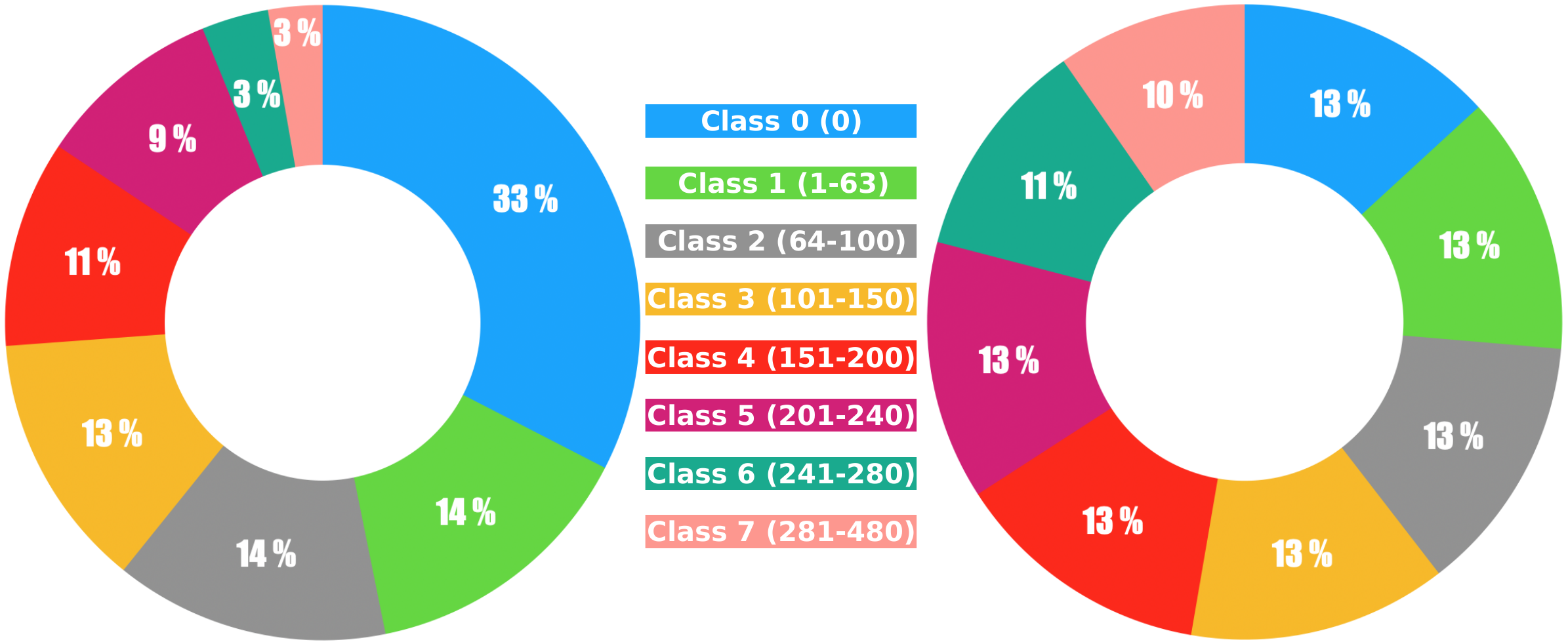}
	\caption{Breakdown of our dataset by partition classes. (a) Unbalanced dataset. (b) Balanced dataset.}
	\label{fig:edge_instances}
\end{figure}

Our dataset includes more than 26 million instances of $64\times64$ block partition, excluding any \gls{ctc} sequence to ensure a fair comparison of our method against state-of-the-art techniques. To analyze the dataset disparity, we first randomly select  2 million instances. Furthermore, to address the data heterogeneity issue, several classes are defined with eight levels of depth partition, from no partition to deep partition. \Figure{\ref{fig:edge_instances}} gives the number of instances in the dataset at each of these eight depth levels, which are defined based on the number of edges activated in the $64\times64$ \glspl{cu}. \Figure{\ref{fig:edge_instances}}-(a) represents the class repetitions through 2 million instances before the dataset balancing. It can be noticed that class 0 without any split contains more than a third of the 2 million $64\times64$ \glspl{cu} instances. In contrast, the last two classes, which illustrate deep partition, are under-represented and must be increased.

\Figure{\ref{fig:edge_instances}}-(b) illustrates the depth distribution of block partitioning over the eight classes after the dataset balancing.
1,200,000 instances ($64\times64$ block) were selected at each \gls{qp} value with a homogeneous representation over the eight classes.
The two last classes are slightly under-represented as highly deep partitions are derived by the encoder only for extremely complex blocks encoded at low \gls{qp} values.


The hard representation is composed of sub-datasets separated by the different \gls{cu} sizes. These sub-datasets are also balanced to enhance their representation.
Indeed, we individually balanced each sub-dataset over the available splits and the \glspl{qp}. For instance, the $4\times16$ \gls{cu} size dataset is balanced between the proportion of \gls{bt}H, \gls{tt}H and no split but also among the 4 \glspl{qp}.

\section{Training process and experimental setup}
\label{sec:training_proc}

\subsection{\gls{cnn} model}

The \gls{cnn} is trained from scratch with the proposed dataset described in Section~\ref{sec:dataset}, relying on the Keras framework~\cite{chollet_et_al_keras_2015} running on top of the Tensorflow module~\cite{abadi_et_al_tensorflow_nodate}.
The weights $\theta$ of the \gls{cnn} are updated at each batch iteration with the ADAM stochastic gradient descent optimizer~\cite{kingma_adam_2017}.  
The loss function is defined to optimize those weights by minimizing the mean squared error between the predicted probability vector $\tilde{\mathbf{p}}$ and the corresponding ground truth vector $\mathbf{p}$ as follows:
\begin{equation}
\label{eq:loss}
\mathcal{L}_{cnn} =  || \mathbf{p} - \tilde{\mathbf{p}}||^2_2.
\end{equation}
The batch size is set to 128 instances ($64\times64$ \glspl{cu}) and the learning rate is equal to $10^{-3}$.  
The training is performed on a hundred epochs with a random shuffle of the dataset at each epoch. The \gls{cnn} training was carried out on a RTX 2080 Ti \gls{gpu}.

\subsection{\gls{dt} \gls{lgbm} model}
The \gls{dt} models are implemented under the \gls{lgbm} framework~\cite{ke_lightgbm_2017} version 2.3.1. This latter is a gradient boosting framework based on a decision tree developed by Microsoft. 
\gls{lgbm} has many advantages, such as low memory usage, the capacity to handle large-scale data, and low inference computational time. This last advantage is essential for our problem as the inference is carried-out at each \gls{cu} level.

\gls{lgbm} is a \gls{dt} method that sums the predictions of all the trees to reach high accuracy. The trees are optimized in a stage-wise way by adding or updating a new tree based on the error of the whole ensemble learned so far. For \gls{dt} classification, the used cross-entropy loss function is defined as follows:
\begin{equation}
\label{eq:loss_multi}
\mathcal{L}_{dt} = - \sum_{i=1}^{6}\mathbf{c}_{i}\log{\tilde{\mathbf{c}}_{i}},
\end{equation}
\noindent where $c_{i}$ is the vector of ground truth split probabilities and $\tilde{c}_{i}$ is the vector of split probabilities predicted by the model.

\subsection{Evaluation procedure and implementation details}
\label{subsec:expe_setup}
All experiments are conducted with the \acrfull{vtm} version 10.2 in \gls{ai} coding configuration.  We consider test video sequences defined in the \gls{vvc} \acrfull{ctc}~\cite{bossen_jvet_2019}. 
The \gls{ctc} video sequences are separated into seven classes as follows: A1 ($3840\times2160$), A2 ($3840\times2160$), B ($1920\times1080$), C ($832\times480$), D ($416\times240$), E ($1280\times720$), and F ($832\times480$ to $1920\times1080$). 
These video sequences are encoded at four \gls{qp} values: 22, 27, 32, and 37. 

The proposed solution is assessed in terms of coding efficiency and computational complexity. The coding efficiency is measured with the \gls{bd-br} metric~\cite{bjontegaard_calculation_2001} that computes the bitrate loss over four \glspl{qp} in percentage with respect to the anchor (i.e., \gls{vtm}10.2) for the same \gls{psnr} objective quality.  The \gls{bd-br} is calculated across the three components, Y, U, and V.
The computational complexity reduction compared to the anchor is assessed by computing the \gls{det} as follows:
\begin{equation}
\label{eq:deltatime}
\Delta ET = \frac{1}{4}\sum_{\gls{qp}_i \in \{22, 27, 32, 37 \}}\frac{T_{R}(\gls{qp}_i) - T_C(\gls{qp}_i)}{T_{R}(\gls{qp}_i)},
\end{equation}
\noindent where $T_{R}$ is the reference encoding time of the \gls{vtm} anchor, and $T_{C}$ is the encoding time of the \gls{vtm} with the proposed complexity reduction method.

Our complexity reduction technique was  integrated into the \gls{vtm}10.2 encoder, which is developed in C++ programming language. 
The \gls{cnn} is built and trained with Python under the Keras framework, and then the model is converted to C++ code with the frugally deep framework~\cite{hermann_frugally_2018}. 
The \gls{dt} models are also trained in Python, and then converted to C++ with the \gls{lgbm} framework~\cite{ke_lightgbm_2017}. 

All encoding operations were carried out sequentially on an Intel Xeon E5-2603 v4 processor running at 1.70 GHz under Ubuntu 16.04.5 operating system (OS).

\section{Experimental results}
\label{sec:expe_res}
In this section, the performance of the proposed method is assessed and analysed. The \gls{cnn} performance is analysed through its accuracy and \gls{roc} curves. The multi-class \gls{dt} classifiers' accuracy is presented through its top-N accuracy.
The complexity reduction proposed method is then assessed in terms of both computational complexity reduction and \gls{bd-br} loss compared to the \gls{vtm}. Multiple configurations of our method are explored depending on the tested top-N \gls{dt} \gls{lgbm} output splits.
Several existing techniques have investigated the complexity reduction of  the new \gls{mtt} partitioning.  We compare our proposed solution with four best-performing state-of-the-art methods, including solutions proposed by Saldanha {\it et al.}~\cite{saldanha_configurable_2021}, Chen {\it et al.}~\cite{chen_fast_2020}, Yang {\it et al.}~\cite{yang_low_2019}, and Li {\it et al.}~\cite{li_deepqtmt_2020}. Finally, we assess the inference overheads of both \gls{cnn} and \gls{dt} models.

\subsection{\gls{cnn} performance}

\begin{figure}[]
	\centering
	\includegraphics[width=1\linewidth]{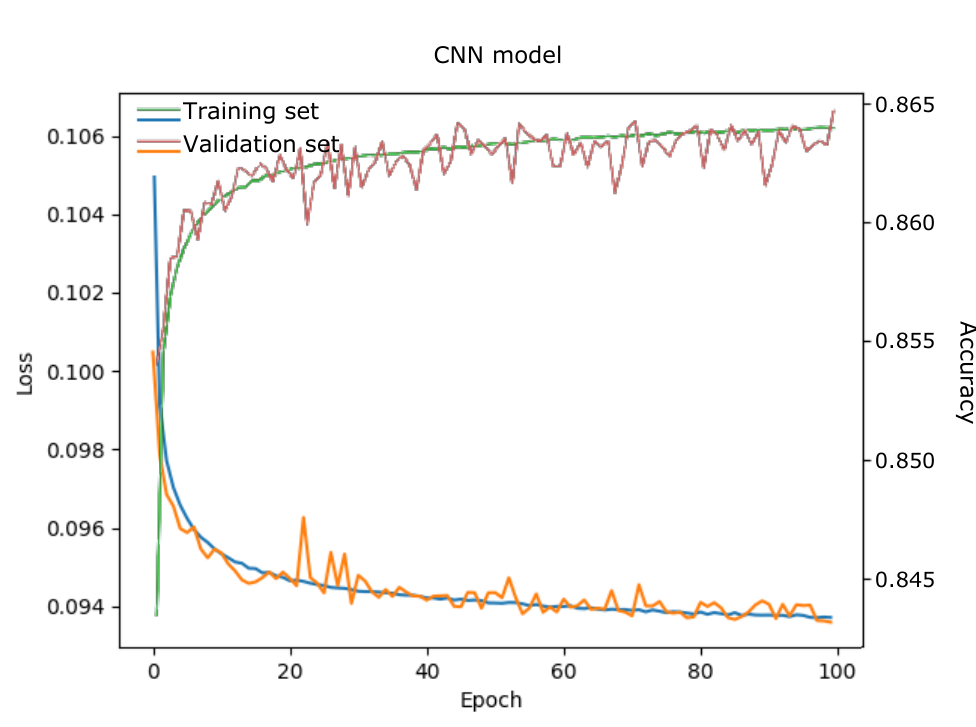}
	\caption{Decreasing loss (in X) and increasing accuracy (in Y) as a function of epoch for the training and validation set.}
	\label{fig:loss_acc_cnn}
\end{figure}

\Figure{\ref{fig:loss_acc_cnn}} shows the loss and accuracy of the \gls{cnn} model versus the training epochs on both training and validation datasets. The blue and orange curves correspond to the losses computed by \Equation{\eqref{eq:loss}} over a hundred training epochs. The green and red curves represent the binary accuracy computed after a shrinkage with a fixed threshold of 0.5. The accuracy is computed between the ground truth and the predicted vector of probabilities. The two curves in blue and green are from the training set, and those in orange and red are from the validation set. The goal of the training is to update the model weights to minimize the loss at each iteration. We notice that the loss curves decrease from 0.105 to 0.094 over 100 epochs for both training and validation sets. The validation curve follows the training curve, which means the model generalizes well on the validation set. The model accuracy reaches more than 86\% of true prediction, i.e., 86\% of the estimated probabilities with a threshold of 0.5 are equal to the ground truth.

\begin{figure}[]
	\centering
	\includegraphics[width=1\linewidth]{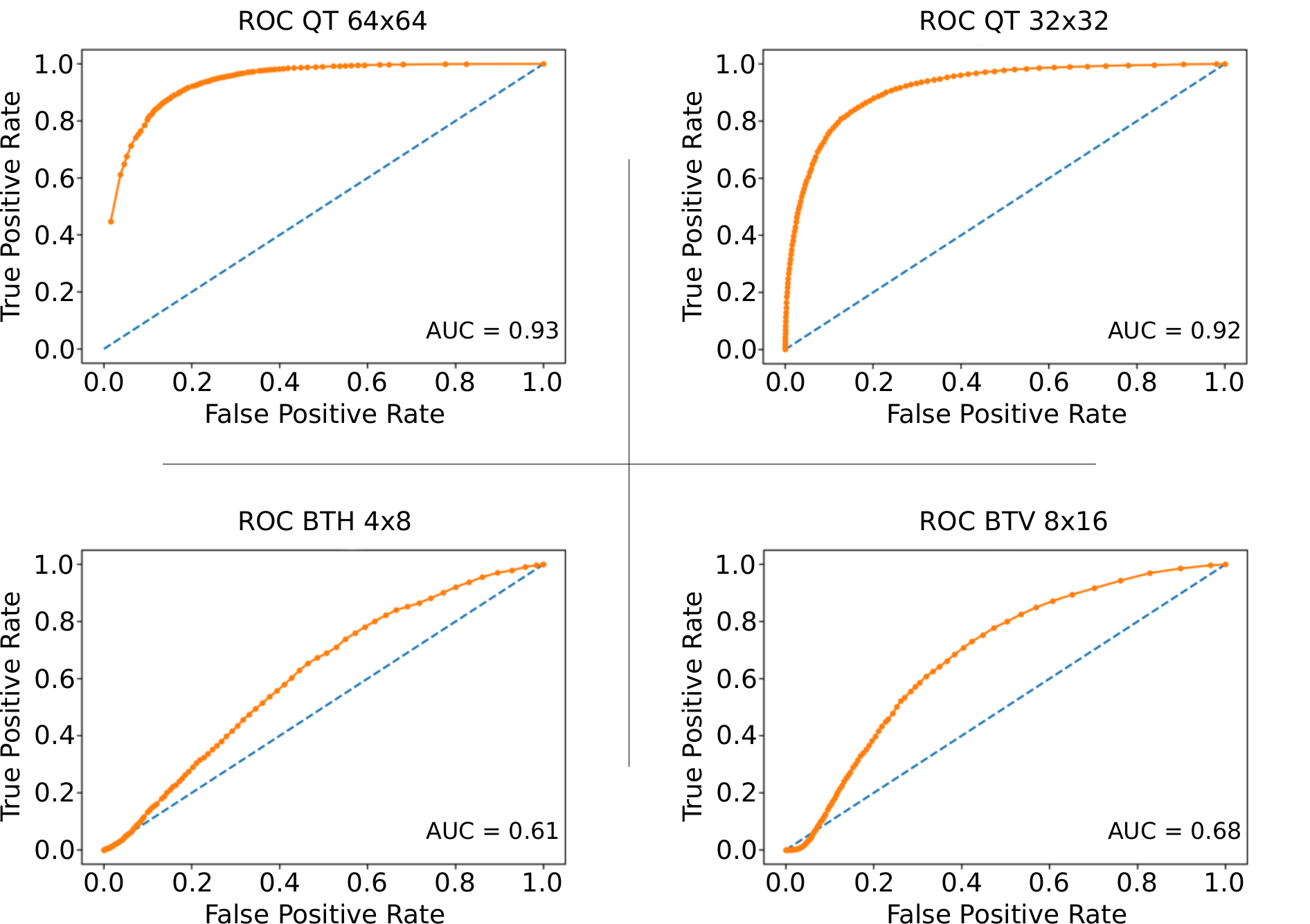}
	\caption{Several \gls{cnn} \glspl{roc} for different splits and sizes on the \gls{ctc} video sequences.}
	\label{fig:roc_cnn}
\end{figure}

The \gls{cnn} prediction performance is also analyzed under \gls{vtm} with the \gls{roc} curves. The \gls{roc} curves represent the true positive rate versus the false positive rate. The split probabilities required to plot these \glspl{roc} are determined by averaging all probabilities at the exact spatial position of the split. \Figure{\ref{fig:roc_cnn}} presents the \gls{roc} curves of \gls{qt}, \gls{bt}H and \gls{bt}V splits at different \gls{cu} sizes computed on the \gls{ctc} sequences. The \gls{qt} \gls{roc} curves show that the average probability is able to reach 0.8 of true positive for approximately false positive rate of 0.1, for both $64\times64$ and $32\times32$ \gls{cu} sizes. Instead, the \gls{bt} \gls{roc} curves are closer to the random guess curve, which is the diagonal one in blue. Indeed, the \gls{cnn} pays more attention to those probabilities. Moreover, the high \gls{cu} sizes more easily determine the split choice as it concerns many pixels. The \gls{roc} area under the curve values are given as a single score to compare the results of the graphs. As shown by the curves, the \gls{roc} area under the curve confirms the results by reaching scores higher than 0.9 for the \gls{qt} and less than 0.7 for the \gls{bt} splits.

\begin{figure*}[]
	\centering
	\includegraphics[width=0.9\linewidth]{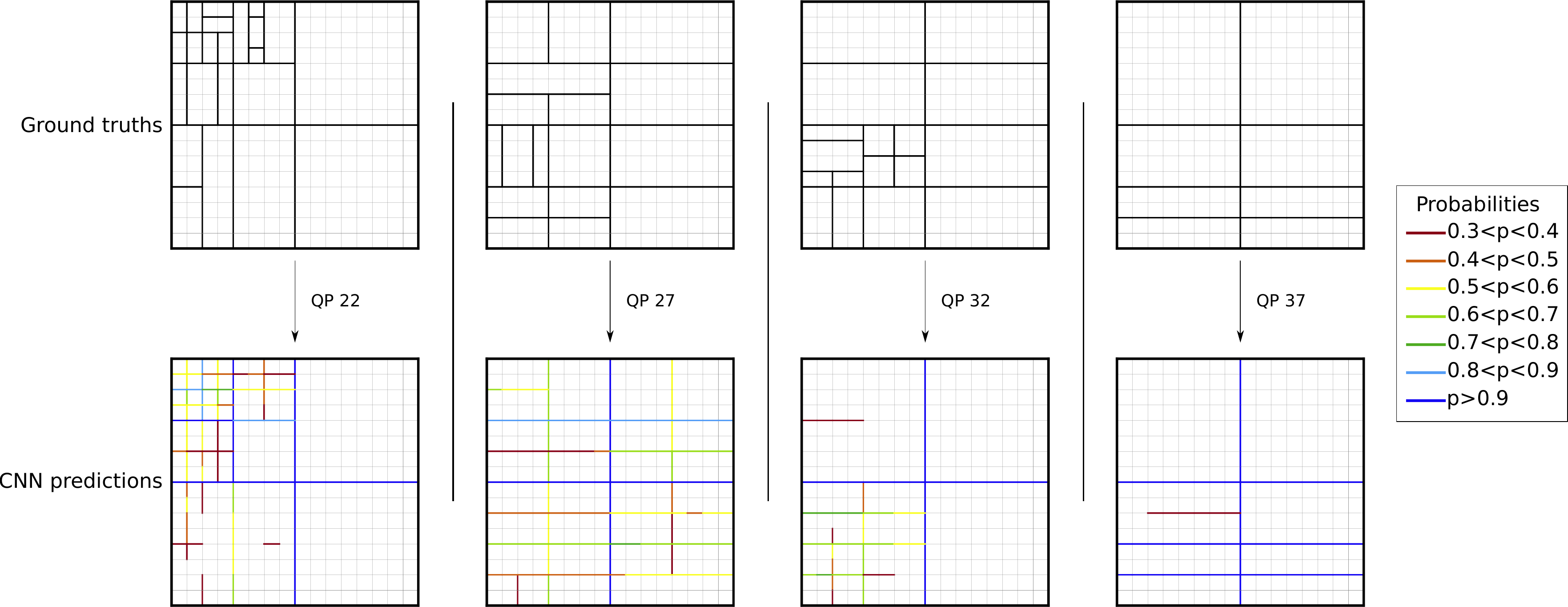}
	\caption{Ground truth partitions on top and their corresponding \gls{cnn} predictions on the bottom for the sequence MarketPlace with \glspl{qp} 22, 27, 32 and 37.}
	\label{fig:cnn_display}
\end{figure*}

Visual illustrations of $64\times64$ \gls{cu} partitions from ground truth and \gls{cnn} prediction are given on the first and second rows of \Figure{\ref{fig:cnn_display}}, respectively.   On the top row, the optimal partitions, derived by the \gls{vtm} anchor, are the ground truth. On the bottom row, the partitions predicted by the \gls{cnn} are displayed with color codes based on their probabilities. The color code varies from red to blue with a probability ranging between 0.3 and 1, and the gray color represents probabilities below 0.3. The partitions for \gls{qp} 37 is shallow with a maximum of three depths and two \glspl{cu} of size $32\times32$.
Its \gls{cnn} prediction is relevant, with each edge defined as a split with a probability higher than 0.9. For the other edges that are not defined as a split, the probability values are lower than 0.3 except for six edges predicted with probabilities between 0.3 and 0.4. Compared to the \gls{qp} 37 partitions, the \gls{qp} 27 configuration is partitioned deeper and is harder to predict efficiently. The figure shows that each edge determined as a split has a probability higher than 0.3 except the \gls{tt} split, and some \glspl{cu} are predicted deeper.

The \gls{cnn} output is not directly used in the \gls{vtm}. It is considered as spatial features and used as an input for the \gls{dt} \gls{lgbm} models. The \gls{dt} \gls{lgbm} will benefit from all the probabilities that impact the partition.

\subsection{\gls{dt} \gls{lgbm} performance}
The second part of the proposed solution includes \gls{dt} \gls{lgbm} models that take advantage of all spatial features predicted by the \gls{cnn} to derive split probabilities.
As presented in \Section{\ref{sec:dataset}}, multiple models are trained to handle the different \gls{cu} sizes. 
The input element range starts from one probability plus the \gls{qp} for $4\times8$ or $8\times4$ \gls{cu} sizes to 480 probabilities plus the \gls{qp} for $64\times64$ \gls{cu} size.
The output is a vector of six classes with \gls{qt}, the two \glspl{bt}, the two \glspl{tt} and no split. For \glspl{cu} with fewer splits available due to \gls{vtm} restrictions, a mask is applied on the predicted vector to restrict the unsuited splits.

\begin{table}[]
	\centering
	\caption{Performance the \gls{dt} \gls{lgbm} models with their defined size and number of output through top-1, top-2 and top-3 accuracy computed on the validation set.}
	\label{tab:ml_res}
	\begin{tabular}{c|c|c|c|c|c}
		\toprule
		Width & Height & \#classes & Top-1 acc. & Top-2 acc. & Top-3 acc.  \\ 	\midrule
		64    & 64     & 2       & 91.69\%              & -                  & -                  \\ 
		32    & 32     & 6       & 59.94\%              & 78.38\%              & 88.87\%              \\ 
		32    & 16     & 5       & 58.50\%              & 77.96\%              & 89.85\%              \\ 
		16    & 32     & 5       & 56.55\%              & 77.48\%              & 89.24\%              \\ 
		32    & 8      & 4       & 54.80\%              & 80.22\%              & 94.11\%              \\ 
		8     & 32     & 4       & 54.91\%              & 80.45\%              & 94.40\%             \\ 
		32    & 4      & 3       & 66.64\%              & 87.80\%              & -                  \\ 
		4     & 32     & 3       & 68.80\%              & 87.38\%             & -                  \\ 
		16    & 16     & 6       & 50.95\%              & 71.27\%              & 84.60\%              \\ 
		16    & 8      & 4       & 62.89\%              & 82.74\%              & 94.05\%              \\ 
		8     & 16     & 4       & 62.36\%              & 83.25\%              & 94.39\%              \\ 
		16    & 4      & 3       & 68.96\%              & 90.12\%              & -                  \\ 
		4     & 16     & 3       & 68.95\%              & 88.54\%             & -                  \\ 
		8     & 8      & 3       & 81.46\%              & 93.05\%              & -                  \\ 
		8     & 4      & 2       & 82.16\%              & -                  & -                  \\ 
		4     & 8      & 2       & 86.26\%              & -                  & -                  \\ \midrule 
		\multicolumn{2}{c|}{\multirow{6}{*}{Average}} & 2 & 86.70\% & - & - \\ 
		\multicolumn{2}{c|}{} & 3 & 70.96\% & 89.38\% & - \\ 
		\multicolumn{2}{c|}{} & 4 & 58.74\% & 81.67\% & 94.24\% \\ 
		\multicolumn{2}{c|}{} & 5 & 57.53\% & 77.72\% & 89.55\% \\ 
		\multicolumn{2}{c|}{} & 6 & 55.45\% & 74.83\% & 86.74\% \\ 
		\multicolumn{2}{c|}{} & - & 67.24\% & 82.97\% & 91.19\% \\ \bottomrule
	\end{tabular}
\end{table}

\Table{\ref{tab:ml_res}} presents the accuracy of the \gls{dt} \gls{lgbm} models on the \gls{ctc} dataset. Three values are reported to analyse the \gls{dt} \gls{lgbm} results based on the top-N accuracy with $N \in \{1, \, 2, \,3 \}$. Top-N accuracy is a metric that measures how often the correct class falls in the top-N highest predicted probabilities. 
Top-1 accuracy reaches, on average, 67.24\% of correct predictions.
Based on the number of available splits, the prediction accuracy is different, decreasing from  86.7\% to 55.45\% for 2 and 6 classes, respectively. The highest top-1 accuracy is achieved with the $64\times64$ block size binary classification between \gls{qt} and no split with 91.69\%. All \gls{dt} \gls{lgbm} models have more than 50\% top-1 accuracy even with multi-class classification.
The top-2 accuracy is given for models with at least three classes, which reaches, on average, 82.97\%.
The lowest accuracy is 71.27\% obtained with the smallest block size available for six classes decision, i.e., $16\times16$.
Finally, top-3 accuracy achieves 91.19\% accuracy on average over the model with at least four classes.
By skipping half of the available splits, the top-3 configuration achieves 86.74\% accuracy for six classes.
This accuracy performance reduces the complexity by a factor of two, with high confidence in predicting the right split.
These results exhibit that the accuracy of \gls{dt} models with three classes depends on the available splits. Higher accuracies are reached for the $8\times8$ block size model in top-1 and top-2 by at least +12\% and +3\%, respectively, compared with other models with three classes.
In addition to the no split mode, two splits are in competition: either \gls{bt} and \gls{tt} in the same direction or \gls{bt} horizontal and vertical. As shown by the results, the direction of a split is easier to predict than the difference between \gls{bt} and \gls{tt} in the same direction.

\subsection{Complexity reduction performance under the \gls{vtm}}

The two-stage proposed model is integrated in the \gls{vtm}10.2 encoder configured in \gls{ai} setting. For a fair comparison, the reference used to compute the \gls{bd-br} loss and the complexity reduction is the classical \gls{vtm} encoder, including multiple native speed-up techniques~\cite{wieckowski_fast_2019} for the tree partitioning process. These speed-up techniques significantly reduce the execution time with a slight \gls{bd-br} degradation compared with an exhaustive \gls{rdo} process. Thus, these experiments exhibit the gain provided by our approach compared with the common configuration of the \gls{vtm} encoder.

We compare our method to the best-performing state-of-the-art methods in complexity reduction and \gls{bd-br} loss. These state-of-the-art methods rely on different \gls{vtm} versions; meanwhile, the tree partitioning tool has not significantly changed during the standardization.

\begin{table*}[]
	\centering
	\caption{ \gls{det}$\uparrow$ and \gls{bd-br}$\downarrow$ performance of the proposed solution in comparison with state of the art techniques in \gls{ai} coding configuration.}
	\label{tab:vtm_res}
	\begin{adjustbox}{max width=1\linewidth}
		\begin{tabular}{l|l|cc|cc|cc|cc|cc|}
		\toprule
			\multirow{2}{*}{Class} & \multirow{2}{*}{Sequence} & \multicolumn{2}{c|}{Saldanha {\it et al.}~\cite{saldanha_configurable_2021}, \gls{vtm}10.0} & \multicolumn{2}{c|}{Chen {\it et al.}~\cite{chen_fast_2020}, \gls{vtm}4.0} & \multicolumn{2}{c|}{Li {\it et al.}~\cite{li_deepqtmt_2020}, \gls{vtm}7.0} & \multicolumn{2}{c|}{Ours top-3, \gls{vtm}10.2} & \multicolumn{2}{c}{Ours top-2, \gls{vtm}10.2} \\ \cmidrule(lr){3-12} 
			& & \gls{bd-br} & \gls{det} & \gls{bd-br} & \gls{det} & \gls{bd-br} & \gls{det} & \gls{bd-br} & \gls{det} & \gls{bd-br} & \multicolumn{1}{c}{\gls{det}} \\ \midrule
			
			\multicolumn{1}{c|}{\multirow{3}{*}{Class A1}} & Tango2 & 0.71\% & 53.1\% & 1.38\% & 48.7\% & 1.52\% & 40.8\% & 0.50\% & 48.5\% & 1.48\% & \multicolumn{1}{c}{80.9\%} \\ 

			\multicolumn{1}{c|}{} & FoodMarket4 & 0.69\% & 46.5\% & 0.84\% & 30.3\% & 1.26\% & 42.2\% & 0.46\% & 46.7\% & 1.4\% & \multicolumn{1}{c}{67.5\%} \\ 

			\multicolumn{1}{c|}{} & Campfire & 0.83\% & 43.8\% & 1.29\% & 49.1\% & 2.02\% & 48.7\% & 0.72\% & 48.2\% & 2.02\% & \multicolumn{1}{c}{67.4\%} \\ \midrule

			\multicolumn{1}{c|}{} & Average & 0.74\% & 47.8\% & 1.17\% & 42.7\% & 1.6\% & 43.9\% & 0.56\% & 47.8\% & 1.63\% & \multicolumn{1}{c}{71.9\%} \\ \midrule 

			\multicolumn{1}{c|}{\multirow{3}{*}{Class A2}} & CatRobot1 & 0.9\% & 44.1\% & 1.99\% & 50.4\% & 2.16\% & 45.4\% & 0.95\% & 46.6\% & 2.56\% & \multicolumn{1}{c}{62.9\%} \\ 

			\multicolumn{1}{c|}{} & DaylightRoad2 & 1.1\% & 53.7\% & 2\% & 54.1\% & 1.16\% & 49.4\% & 1.00\% & 52.1\% & 2.47\% & \multicolumn{1}{c}{73.7\%} \\ 

			\multicolumn{1}{c|}{} & ParkRunning3 & 0.45\% & 41.4\% & 0.8\% & 40.6\% & 1.15\% & 41.7\% & 0.32\% & 43.3\% & 0.91\% & \multicolumn{1}{c}{60.9\%} \\ \midrule

			\multicolumn{1}{c|}{} & Average & 0.82\% & 46.4\% & 1.6\% & 48.4\% & 1.49\% & 45.5\% & 0.76\% & 47.4\% & 1.98\% & \multicolumn{1}{c}{65.9\%} \\ \midrule 

			\multicolumn{1}{c|}{\multirow{5}{*}{Class B}} & MarketPlace & 0.6\% & 54.5\% & - & - & 0.8\% & 46.6\% & 0.46\% & 55.0\% & 1.29\% & \multicolumn{1}{c}{75.7\%} \\ 

			\multicolumn{1}{c|}{} & RitualDance & 1.09\% & 53.5\% & - & - & 1.07\% & 44.9\% & 0.80\% & 51.4\% & 2.41\% & \multicolumn{1}{c}{73.4\%} \\ 

			\multicolumn{1}{c|}{} & Cactus & 1.04\% & 50\% & 1.73\% & 49.8\% & 1.12\% & 49.3\% & 0.84\% & 49.5\% & 2.6\% & \multicolumn{1}{c}{72.8\%} \\ 

			\multicolumn{1}{c|}{} & BasketballDrive & 1.26\% & 57.1\% & 1.54\% & 50.1\% & 1.64\% & 52\% & 0.88\% & 52.1\% & 2.51\% & \multicolumn{1}{c}{76.0\%} \\ 

			\multicolumn{1}{c|}{} & BQTerrace & 1.11\% & 48.3\% & 1.4\% & 56\% & 1.11\% & 45.6\% & 1.02\% & 47.0\% & 2.76\% & \multicolumn{1}{c}{75.1\%} \\ \midrule

			\multicolumn{1}{c|}{} & Average & 1.02\% & 52.7\% & 1.56\% & 52\% & 1.15\% & 47.7\% & 0.80\% & 51.0\% & 2.31\% & \multicolumn{1}{c}{74.6\%} \\ \midrule 

			\multicolumn{1}{c|}{\multirow{4}{*}{Class C}} & RaceHorses & 0.75\% & 46.9\% & 1.35\% & 50.6\% & 0.96\% & 46.5\% & 0.61\% & 45.3\% & 1.97\% & \multicolumn{1}{c}{69.2\%} \\ 

			\multicolumn{1}{c|}{} & BQMall & 1.4\% & 51\% & 2.12\% & 58.9\% & 1.17\% & 49.8\% & 0.94\% & 46.8\% & 3.08\% & \multicolumn{1}{c}{71.1\%} \\ 

			\multicolumn{1}{c|}{} & PartyScene & 0.77\% & 48.3\% & 1.01\% & 51\% & 0.61\% & 45.2\% & 0.54\% & 43.7\% & 2.16\% & \multicolumn{1}{c}{69.3\%} \\ 

			\multicolumn{1}{c|}{} & BasketballDrill & 1.52\% & 40.3\% & 2.05\% & 54.8\% & 1.63\% & 39.3\% & 1.48\% & 44.7\% & 4.69\% & \multicolumn{1}{c}{67.7\%} \\ \midrule

			\multicolumn{1}{c|}{} & Average & 1.11\% & 46.6\% & 1.63\% & 53.8\% & 1.09\% & 45.2\% & 0.89\% & 45.1\% & 2.97\% & \multicolumn{1}{c}{69.3\%} \\ \midrule 

			\multicolumn{1}{c|}{\multirow{4}{*}{Class D}} & RaceHorses & 0.72\% & 45\% & 1.28\% & 54.7\% & 1.2\% & 41.6\% & 0.60\% & 43.9\% & 2.29\% & \multicolumn{1}{c}{66.5\%} \\ 

			\multicolumn{1}{c|}{} & BQSquare & 0.57\% & 40.7\% & 0.75\% & 52.8\% & 0.74\% & 44.5\% & 0.55\% & 44.0\% & 2.43\% & \multicolumn{1}{c}{70.0\%} \\ 

			\multicolumn{1}{c|}{} & BlowingBubbles & 0.82\% & 43.7\% & 1.4\% & 54.9\% & 0.92\% & 41.6\% & 0.60\% & 40.2\% & 2.37\% & \multicolumn{1}{c}{64.7\%} \\ 

			\multicolumn{1}{c|}{} & BasketballPass & 1.32\% & 49.3\% & 1.77\% & 53.1\% & 1.41\% & 44.5\% & 0.84\% & 44.5\% & 2.86\% & \multicolumn{1}{c}{67.2\%} \\ \midrule

			\multicolumn{1}{c|}{} & Average & 0.86\% & 44.7\% & 1.3\% & 53.9\% & 1.07\% & 43.1\% & 0.64\% & 43.1\% & 2.49\% & \multicolumn{1}{c}{67.1\%} \\ \midrule 

			\multicolumn{1}{c|}{\multirow{3}{*}{Class E}} & FourPeople & 1.71\% & 54.1\% & 2.71\% & 56.3\% & 1.33\% & 49.9\% & 1.10\% & 49.9\% & 3.46\% & \multicolumn{1}{c}{73.1\%} \\ 

			\multicolumn{1}{c|}{} & Johnny & 1.65\% & 55.3\% & 2.77\% & 55.8\% & 2.33\% & 48.2\% & 1.34\% & 50.5\% & 3.71\% & \multicolumn{1}{c}{72.9\%} \\ 

			\multicolumn{1}{c|}{} & KristenAndSara & 1.26\% & 53\% & 2.17\% & 52.8\% & 1.76\% & 50.5\% & 0.97\% & 49.0\% & 3.32\% & \multicolumn{1}{c}{71.8\%} \\ \midrule

			\multicolumn{1}{c|}{} & Average & 1.54\% & 54.1\% & 2.55\% & 55\% & 1.81\% & 49.5\% & 1.14\% & 49.8\% & 3.5\% & \multicolumn{1}{c}{72.6\%} \\ \midrule

	\multicolumn{2}{c|}{ \textbf{Average}} 		 & 1.01\% & 48.8\% & 1.62\% & 51.2\% & 1.32\% & 45.8\% & 0.79\% & 47.4\% & 2.49\% & \multicolumn{1}{c}{70.4\%} \\ \midrule 

			\multicolumn{1}{c|}{\multirow{4}{*}{Class F}} & ArenaOfValor & - & - & - & - & - & - & 1.00\% & 44.9\% & 3.28\% & \multicolumn{1}{c}{66.2\%} \\ 

			\multicolumn{1}{c|}{} & BasketballDrillText & - & - & 2.09\% & 56.3\% & - & - & 1.57\% & 43.3\% & 4.69\% & \multicolumn{1}{c}{67.7\%} \\ 

			\multicolumn{1}{c|}{} & SlideEditing & - & - & 0.52\% & 45.4\% & - & - & 0.95\% & 46.1\% & 4.14\% & \multicolumn{1}{c}{69.7\%} \\ 

			\multicolumn{1}{c|}{} & SlideShow & - & - & 2.11\% & 45.8\% & - & - & 1.39\% & 44.4\% & 4.56\% & \multicolumn{1}{c}{68.0\%} \\ \midrule

			\multicolumn{1}{c|}{} & Average & - & - & 1.57\% & 49.2\% & - & - & 1.23\% & 44.7\% & 4.17\% & \multicolumn{1}{c}{67.9\%} \\ \bottomrule
		\end{tabular}
	\end{adjustbox}
\end{table*}

\Table{\ref{tab:vtm_res}} presents the \gls{bd-br} and complexity reduction performance of our method compared with the state-of-the-art techniques proposed by Saldanha {\it et al.}~\cite{saldanha_configurable_2021}, Chen {\it et al.}~\cite{chen_fast_2020}, and Li {\it et al.}~\cite{li_deepqtmt_2020}.
In this table, two configurations are presented based on the top-2 and top-3 configurations. The results are illustrated for the \gls{ctc} sequences from class A1 to class F.
The average is given for each class independently, and all the sequences from class A1 to class E. However, class F is not considered in the average computation as it includes specific video sequences such as screen content.

On average, our top-3 configuration through all sequences reaches 0.79\% \gls{bd-br} loss for a complexity reduction of 47.4\%. Compared with Li {\it et al.}~\cite{li_deepqtmt_2020} method, our solution achieves better performance in both \gls{bd-br} and complexity reduction. Chen {\it et al.}~\cite{chen_fast_2020} solution reduces 4.2\% more complexity but at the expense of a significant increase in \gls{bd-br} loss of 0.83\% compared with our top-3 configuration. This low gain in the complexity reduction is not relevant compared with the loss in \gls{bd-br}, which is almost doubled. The method proposed by Saldanha {\it et al.}~\cite{saldanha_configurable_2021} enhances the complexity reduction by 1.5\% but for a 0.22\% more \gls{bd-br} loss. The second configuration with top-2 achieves, on average, 2.49\% \gls{bd-br} loss with 70.4\% of complexity reduction.

High-resolution videos  are the main interest in the \gls{vvc} development. Tackling the complexity reduction of high-resolution videos is particularly important as the encoding time is proportional to the sequence resolution. Classes A and B represent high-resolution sequences with 4K and full HD, respectively. Our top-3 and top-2 configurations are able to reach 47.6\% and 68.9\% complexity reductions for 0.66\% and 1.80\% \gls{bd-br} losses on class A sequences, respectively. In both \gls{bd-br} and complexity reduction metrics our method is better than Chen {\it et al.}~\cite{chen_fast_2020} and Li {\it et al.}~\cite{li_deepqtmt_2020} techniques.
Compared with these two techniques, our top-3 configuration solution achieves, on average, a lower \gls{bd-br} loss of 0.69\% and 0.85\% and a higher computational complexity reduction by 1.5\% and 2.3\%, respectively. Saldanha {\it et al.}~\cite{saldanha_configurable_2021} solution obtains slightly lower complexity reduction with a  a 0.12\% \gls{bd-br} loss compared to our top-3 configuration. Concerning the class B sequences, our top-3 configuration can halve the complexity with 51\% of complexity reduction for a 0.8\% \gls{bd-br} loss.
The closest to our method is Saldanha {\it et al.}~\cite{saldanha_configurable_2021} solution which achieves slightly more complexity reduction score but for an increase of 0.22\% \gls{bd-br} loss.  Our top-2 configuration enables 74.6\% complexity reduction for 2.31\% \gls{bd-br} loss. The highest performance in complexity reduction and \gls{bd-br} loss is achieved for the {\it MarketPlace} sequence. Indeed, the top-3 and top-2 configurations reach 55\% and 75.7\% complexity reductions for 0.46\% and 1.29\% \gls{bd-br} losses, respectively.

For lower resolution classes C to E, the top-3 configuration achieves less than 1\% \gls{bd-br} loss for a maximum of 50.5\% complexity reduction.
Compared to Saldanha {\it et al.}~\cite{saldanha_configurable_2021}, our method has approximately the same complexity reduction but obtains lower \gls{bd-br} loss under classes C and D.
For class E, their solution achieves 4.3\% higher complexity reduction but with an increase of 0.4\% \gls{bd-br} loss.
Li {\it et al.}~\cite{li_deepqtmt_2020} has  a higher \gls{bd-br} loss for lower complexity reduction for classes C and D compared with our top-3 configuration. 
For class E, Li {\it et al.}~\cite{li_deepqtmt_2020} solution achieved the same complexity reduction but at the cost of 0.53\% \gls{bd-br} loss compared with our method.
Chen {\it et al.}~\cite{chen_fast_2020} almost doubled the \gls{bd-br} loss for a complexity reduction increase of 3.8\% compared with our top-3 configuration through low resolution classes.
In the case of the top-2 configuration, the performance is lower on low resolutions than on high-resolution. This approach on low-resolution contents, including classes C, D, and E reaches the complexity of high-resolution contents (classes A1, A2, and B) of 69.66\% for a higher \gls{bd-br} loss of 2.98\% which is higher by approximately 1\% compared with the \gls{bd-br} loss of high resolutions contents (1.97\%).
Low-resolution contents result in deeper partitions, so the more the complexity is reduced, the less space is available to skip splits. Therefore, as the global partition is composed of more splits, the impact on \gls{bd-br} is more significant at high complexity reductions.

Class F has specific sequences with screen contents such as slides or gaming content.
Our method still achieves 44.7\% complexity reduction for 1.23\% \gls{bd-br} loss. The work of Chen {\it et al.}~\cite{chen_fast_2020} is the only one that reported results on class F with three out of four sequences. Our solution performs a slightly lower complexity reduction with a less \gls{bd-br} loss.

\begin{figure}[t]
	\centering
	\includegraphics[width=1\linewidth]{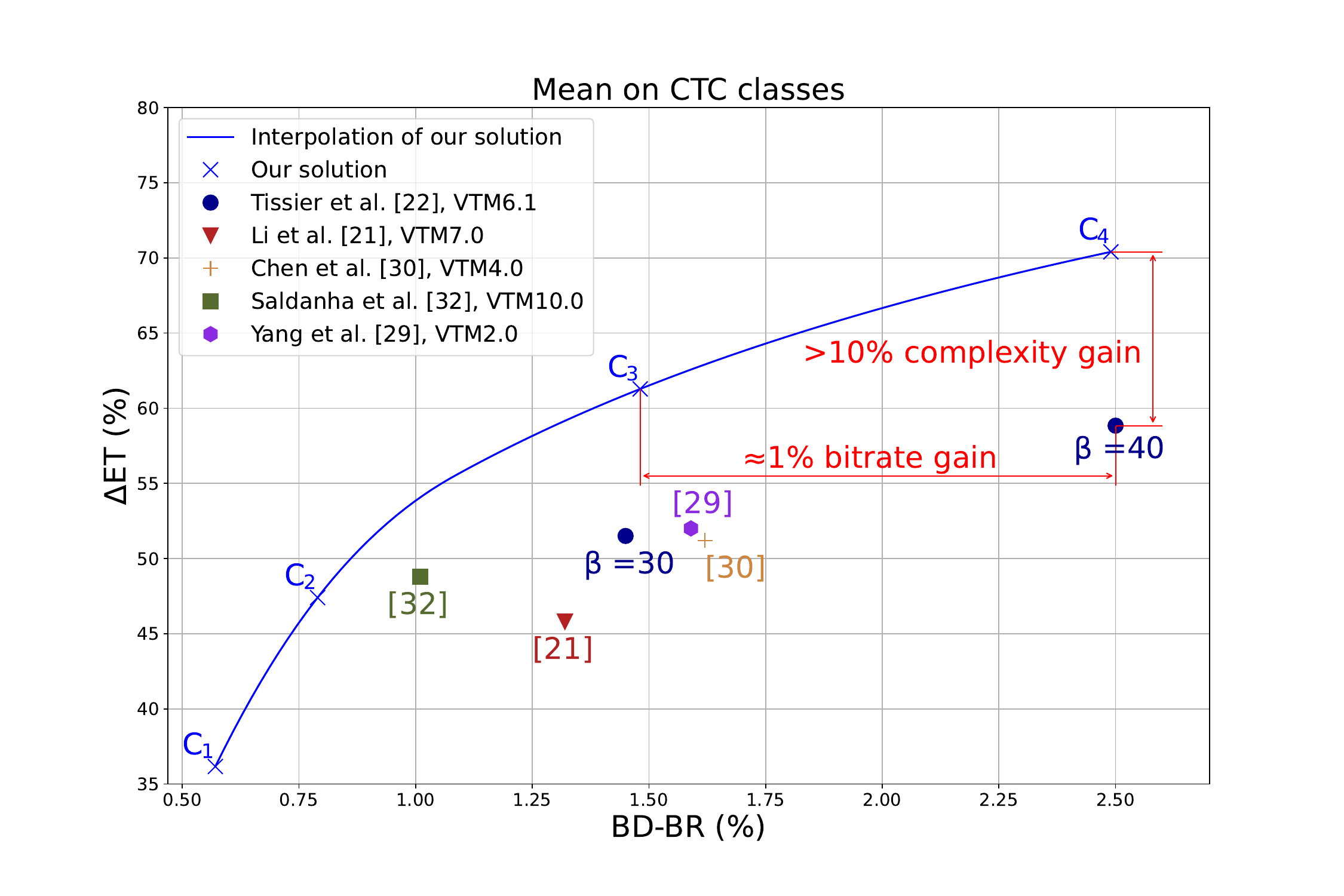}
	\caption{Complexity reduction versus \gls{bd-br} performance comparison between the proposed method and state of the art techniques on \gls{ctc} classes except class F (\gls{ai} configuration). An interpolation cure over our four configurations is plotted in a blue dot line.}
	\label{fig:res_vtm_graph}
\end{figure}

\Figure{\ref{fig:res_vtm_graph}} illustrates the performance of our method in complexity reduction versus \gls{bd-br} jointly with state-of-the-art methods in a 2D plan averaged on the \gls{ctc} classes excluding class F. 
Saldanha {\it et al.}~\cite{saldanha_configurable_2021} proposed different trade-offs depicted in this figure. In addition to the configurations presented in \Table{\ref{tab:vtm_res}}, two new configurations are included. The selection of the top-3 modes for the multi-classification with 6 outputs (classes) and top-4 modes for the other \gls{dt} \gls{lgbm} models is defined as $C_1$. The top-3 configuration is defined as $C_2$. The top-3 for all \gls{dt} \gls{lgbm} models except the multi-classification with 6 outputs for which top-2 is used in configuration $C_3$, and finally, top-2 for all \gls{dt} models is defined as $C_4$.
These configurations reach different trade-off points between complexity reduction and \gls{bd-br} loss, allowing us to draw an interpolation curve over these four configurations.
This interpolation helps us to compare the results since the rate-distortion-complexity trade-off points are not linear.
As explained in \Table{\ref{tab:vtm_res}}, the figure confirms that our solution outperforms the best-performing state-of-the-art techniques. The points representing the state-of-the-art solutions are below our approach's interpolation curve.  

To illustrate the gains brought by introducing \gls{dt} models and dataset balancing, we show in \Figure{\ref{fig:res_vtm_graph}}  two configurations of our previous solution~\cite{tissier_cnn_2020} relying on decision thresholds with $\beta = 30$ and $\beta = 40$. Our $C_3$ and $C_4$ configurations are better than the $\beta = 30$ and $\beta = 40$ solutions with a gain in complexity reduction of approximately 10\% for a similar \gls{bd-br} loss. Compared with $\beta = 40$ which reaches 58.8\% complexity reduction, our solution $C_3$ affords a significant \gls{bd-br} gain of 1.02\%.

\begin{table}
\centering
\caption{Ablation study of the proposed method. Performance in terms of  \gls{det}$\uparrow$, \gls{bd-br}$\downarrow$  and the ratio \gls{det}/\gls{bd-br}$\uparrow$. The baseline model is the \gls{cnn} prediction with a threshold $\beta=30$, baseline + DT includes the DT models for the prediction, and ours uses the dataset balancing. The best ratio is highlighted in bold.}
\label{tab:ablation}
	\begin{adjustbox}{max width=1\linewidth}
\begin{tabular}{l|c|c|c} 
\toprule
Class                    & Baseline ($\beta=30$)~\cite{tissier_cnn_2020}         & Baseline + DT         & Ours ($C_3$)        \\ \midrule
 A1                & 1.55/62.9 (40.58)         & 0.61/47.5 (79.17)          & 0.56/47.8 ({\bf 85.35})         \\
 A2                  & 1.47/60.0 (40.81)         & 0.80/46.4 (58.00)          & 0.76/47.4 ({\bf 62.36})        \\
 B             &1.41/61.1 (43.33)        &  0.85/51.0 (60.00)          & 0.80/51.0 ({\bf 63.75})          \\
 C        & 1.20/37.9 (29.37)          & 0.96/45.7 (47.60)          & 0.89/51.1 ({\bf 57.41})         \\
 D      & 0.83/32.5 (39.15)          & 0.70/43.3 (61.86)         & 0.64/43.1 ({\bf 67.34})         \\
 E             & 2.29/54.4 (23.75)          & 1.23/43.9 (35.69)          & 1.14/49.8 ({\bf 43.68})         \\ \midrule
Ave.        & 1.45/51.1 (35.24)          & 0.86/46.6 (54.18)          & 0.79/47.4 ({\bf 60.00})        \\ \midrule
 F            & 1.61/36.3 (22.54)       & 1.34/24.5 (18.28)          & 1.23/44.7 ({\bf 36.34})         \\
\bottomrule
\end{tabular}
\end{adjustbox}
\end{table}

\subsection{Ablation study}
\Table{\ref{tab:ablation}} shows the contributions of the \gls{dt} \gls{lgbm} models and the dataset balancing  on the top of the baseline model that considers only the \gls{cnn} prediction with a simple threshold $\beta=30$~\cite{tissier_cnn_2020}.   We can notice that using the \gls{dt} \gls{lgbm} models enhance the \gls{det}/\gls{bd-br} ratio on average from 35.24 to 54.18. Furthermore, dataset balancing enables consistent gains, reaching the highest \gls{det}/\gls{bd-br} ratios over all \gls{ctc} classes.

\section{Complexity analysis}
\label{sec:complex_study}
In this section we assess the complexity overhead of the \gls{cnn} and \gls{dt} inferences under the \gls{vtm}.  Then, optimizations are proposed to reduce the complexity of the \gls{ml} prediction models.

\subsection{Complexity overhead}
\Table{\ref{tab:time_cnn_ml}} presents the time spent in the \gls{cnn} and in the \gls{dt} predictions for the $C_2$ configuration in comparison with the \gls{vtm}10.2 anchor encoding time. These values are obtained by computing the ratio between the \gls{cnn} or \gls{dt} time and the \gls{vtm} reference encoding time. The execution time of the \gls{cnn} inference is, on average lower than the execution time of the \gls{dt} model.
Indeed, even if the \gls{cnn} inference is more complex, the \gls{dt} infers at each \gls{cu} size unlike the \gls{cnn} which is carried-out only once for each $64\times64$ \gls{cu}.
The run-time ratio of the \gls{cnn} is higher for the \gls{uhd} classes, primarily class A1. This is caused by the early skip methods integrated into the \gls{vtm} that lead to shallow partition, resulting in a faster encoding process.
Moreover, the results for the \gls{dt} inference time is homogeneous through all the \gls{ctc} classes. On average, the run-time of both the \gls{cnn} and \gls{dt} is under 2\%, taken together; they require less than 3\% of the encoding time.
These results show that the prediction times are negligible, especially since the \gls{cnn} and \gls{dt} models can be optimized, accelerated, or run in parallel with the encoding.
Indeed, the execution time of the \gls{cnn} can be significantly reduced by targeting a \gls{gpu} or on multi-core processors with optimizations, as presented below.

\begin{table}[t]
	\centering
	\caption{Complexity overhead of the \gls{cnn} and \gls{dt} \gls{lgbm} (in \%) for $C_2$ configuration compared to the run time of the \gls{vtm} anchor.}
	\label{tab:time_cnn_ml}
	\begin{tabular}{l|c|c|c}
	\toprule
		& \gls{cnn} coml. overhead  & \gls{dt} coml. overhead  & Total   \\ \midrule
	Class A1	 & 2.7\% & 2.0\% & 4.7\%  \\ 
		Class A2  &  1.2\% & 1.7\% & 2.9\%   \\ 
		Class B   &  1.0\% & 1.7\%    & 2.7\%   \\ 
		Class C & 0.6\% & 1.6\% &  2.2\% \\ 
		Class D & 0.5\% & 1.6\% & 2.1\% \\ 
		Class E & 1.4\% & 1.8\% & 3.2\% \\ 
		Class F &1.2\% & 1.6\% & 2.8\% \\  \midrule
		Average & 1.2\%  & 1.7\% & 2.9\% \\ \bottomrule
	\end{tabular}
\end{table}

\subsection{CNN inference optimisation}
\label{sec:cnn_inf_optim}
\gls{cnn}-based complexity reduction methods have achieved outstanding results for the \gls{vtm} encoder.
However, \gls{cnn} complexity needs to be carefully optimized to minimize the complexity overhead of the prediction.
The proposed \gls{cnn} processes as input a luminance block of size $68\times68$ with ten convolution layers and a final fully connected layer.
The proposed \gls{cnn} model consists of 226 088 training parameters.
The execution time of the \gls{cnn} inference is evaluated on \gls{cpu} and \gls{gpu} platforms as detailed below.

\paragraph{Optimisation on \acrlong{cpu}} 
As presented previously, the frugally-deep framework was considered to generate the C++ source code for the \gls{cnn} inference, which is then integrated into the \gls{vtm}. The proposed solution computed one \gls{cnn} inference for each $64\times64$ block of pixels to predict the partitions. The \gls{cnn} execution time depends on the targeted \gls{cpu}.

The \gls{cnn} inference compiled without specific compilation options and with one thread lasts 153ms for a $64\times64$ block. The inference complexity is also studied under the TensorFlow framework with different \glspl{cpu}.

\Table{\ref{tab:cpu_inference_tf}} lists the different \glspl{cpu} and \glspl{gpu} used to infer with our \gls{cnn} through the TensorFlow framework and provides their respective inference times. The inference time depends mainly on the \gls{cpu} clock rate.
Indeed, with the Xeon W-2125 (8 cores - 4 GHz), the inference time to predict the partition of a $64\times64$ block is 2.13ms.
The slowest \gls{cpu} is the I5-10300H (4 cores - 2.5 GHz) with 3.36ms per inference.
The inference time under the frugally-deep framework is at least $50\times$ higher than under the TensorFlow framework, which includes single instruction multiple data optimizations. 

\paragraph{Optimisation on \acrlong{gpu}}
The \gls{gpu} is more adapted to train and infer \gls{cnn} as most of the computations are matrix based and can be computed in parallel.
Furthermore, the CUDA \gls{api} that manages parallel computing along the cuDNN framework that optimizes standard operation for \gls{cnn} are available to improve the inference execution time.
Different versions of TensorFlow and CUDA are tested, impacting the performance of the \gls{cnn}.
For these experiments, the \gls{gpu} selected is the Nvidia GTX 1650 Max Q.
Under the TensorFlow framework specialized for \gls{gpu} version 2.0.0 and the CUDA version 10.0, the \gls{cnn} infers at $254.65\mu s$ for a $64\times64$ block partition, resulting in a 7.7 \gls{fps} on full HD resolution.  

Optimizing the \gls{cnn} model is proposed to improve the inference execution time. The TensorRT framework proposed by Nvidia defined different optimizations like layer or tensor fusion to optimize the \gls{gpu} memory, bandwidth, and precision refinement by quantizing the \gls{cnn} model.

\Table{\ref{tab:cpu_inference_tf}} details the inference time under the TensorRT framework with the \gls{gpu} Nvidia GTX 1650 Max Q  with different versions of the TensorRT framework with 32-bit and 16-bit floating-point data types. The fastest configuration is the TF-TRT 2.0.0 with FP16, which can predict the partition of the $64\times64$ block at $120.8\mu s$, which leads to 16.2 \gls{fps} on a full \gls{hd} sequence, while with FP32 configuration, the inference reaches $135.17\mu s$. These optimizations show impressive results using a \gls{gpu} with TensorRT. Moreover, a dedicated processor can be used for inference, such as neural processing units proposed by Huawei or tensor processing units offered by Google, to reach even higher performance.
Another option is to change the \gls{cnn} architecture to limit the number of parameters and computations to reduce its inference run time.

\begin{table}[t]
	\centering
	\caption{Inference time under the TensorFlow framework under different \gls{cpu} and \gls{gpu} platforms. The inference time is computed on a 64$\times$64 block , and the frame rate on a full HD resolution video.}
	\label{tab:cpu_inference_tf}
	\begin{adjustbox}{max width=1\linewidth}
		\begin{tabular}{l|l|c|c}
		\toprule
		                      \multicolumn{2}{c|}{Platform}       & Inference time  &  Frame rate (fps) \\ \midrule
				\multirow{4}{*}{ \rotatebox[origin=c]{90}{ \gls{cpu}}}    & Xeon W-2125 (8 cores - 4 GHz)     & 2.13 ms     &   0.92  \\ 
			& Ryzen 5 2600X (6 cores - 3.6 GHz) & 2.53 ms    &  0.78   \\
			& I7-8700 (6 cores - 3.2 GHz)       & 2.66 ms      &  0.74 \\ 
			& I5-10300H (4 cores - 2.5 GHz)     & 3.36  ms & 0.59 \\  
			\midrule
		\multirow{3}{*}{ \rotatebox[origin=c]{90}{ \gls{gpu}}}   	& TF-TRT 2.4.0     & 259.49 $\mu s$  &     7.6   \\ 
			& TF-TRT 2.0.0 (FP32) & 135.17 $\mu s$  &   14.5    \\ 
			& TF-TRT 2.0.0 (FP16)      & 120.80 $\mu s$ &       16.2 \\ \bottomrule
		\end{tabular}
	\end{adjustbox}
\end{table}


\section{Conclusion}
\label{sec:conclu}
In this paper we have proposed a two stage \gls{cnn} and \gls{dt} method to reduce the complexity of the \gls{vtm} encoder in \gls{ai} configuration.
The \gls{cnn} is designed to predict a \gls{cu} partition through a vector of probabilities based on the local activity of pixels in a block. This vector considered as spatial features is then passed as input to the \gls{dt} model that predicts the split probabilities at each \gls{cu} depth. The \gls{dt} method integrated after the \gls{cnn} benefits from all the probabilities inside the computed \gls{cu} instead of taking only the probabilities at the spatial location of the split. Depending on the selected configuration, a top-N probability selection is performed on the \gls{dt} output to skip the unlikely splits.

Our proposed method outperforms state-of-the-art techniques regarding the trade-off between complexity reduction and coding loss. Concerning the top-3 configuration, our proposal assessed on the \gls{vvc} \gls{ctc} sequences enabled on average 47.4\% complexity reduction for a negligible \gls{bd-br} loss of 0.79\%. The top-2 configuration was able to reach a higher complexity reduction of 69.8\% for 2.57\% \gls{bd-br} loss.
The complexity of the \gls{ml} model has been carefully optimized. The \gls{cnn} is able to predict the partition of the $64\times64$ block partition at $120.8\mu s$, resulting in 16.2 \gls{fps} on a full \gls{hd} sequence. These promising results motivated us to extend our method to the \gls{ra} configuration. This extension, investigated as future work, will require taking advantage of the motion flow among adjacent frames.
\section*{Acknowledgments}
This work is supported by the Hubert Curien Partnerships (PHC) Maghreb 2021, No 45988WG (Eco-VVC project), and Région Bretagne through the TRISTRAM collaborative project and the Allocations de Recherche Doctorale (ARED) program.

\newpage

\ifCLASSOPTIONcaptionsoff
\newpage
\fi

\bibliographystyle{IEEEtran}

\bibliography{bibliography}

%








\end{document}

%% file: acronyms.tex
\newacronym{avc}{AVC}{Advanced Video Coding}

\newacronym{ai}{AI}{All Intra}

\newacronym{api}{API}{Application Programming Interface}

\newacronym{awms}{AWMS}{Adaptive Workload Management Scheme}

\newacronym{acpi}{ACPI}{Advanced Configuration and Power Interface}

\newacronym{ate}{ATE}{Absolute Time Error}

\newacronym{aom}{AOM}{Alliance for Open Media}

\newacronym{arc}{ARC}{Adaptive Rate Control}
\newacronym{ann}{ANN}{artificial neural network}
\newacronym{av1}{AV1}{AOMedia Video 1}

\newacronym{alf}{ALF}{Adaptive Loop Filter}

\newacronym{bd-br}{BD-BR}{Bj\o ntegaard Delta Bit Rate}
\newacronym{bd-psnr}{BD-\acrshort{psnr}}{Bj\o ntegaard Delta \acrshort{psnr}}
\newacronym{ber}{BER}{Bit Error Rate}
\newacronym{bu}{BU}{Bottom-up}

\newacronym{bo}{BO}{Band Offset}

\newacronym{bpr}{BPR}{Bi-Prediction Refinement}
\newacronym{bt}{BT}{Binary-Tree}
\newacronym{ctu}{CTU}{Coding Tree Unit}
\newacronym{cu}{CU}{Coding Unit}
\newacronym{ctb}{CTB}{Coding Tree Block}
\newacronym{cb}{CB}{Coding Block}

\newacronym{cpb}{CPB}{Current Picture Buffer}

\newacronym{cpu}{CPU}{Central Processing Unit}

\newacronym{ct}{CT}{Coding-tree partitioning}

\newacronym{cdc}{CDC}{Constrain the Docile \glspl{ctu}}

\newacronym{cdf}{CDF}{Cumulative Distribution Function}
\newacronym{cdm}{\textsc{Cdm}}{\gls{ctu} Depth Map}

\newacronym{cabac}{CABAC}{Context-Adaptive Binary Arithmetic Coding}

\newacronym{cbf}{CBF}{Coded Block Flag}

\newacronym{ccs}{CCS}{Complexity Control Scheme}

\newacronym{ctde}{CTDE}{Coding Tree Depth Estimation}
\newacronym{ccupu}{CCUPU}{Constrained Coding Units and Prediction Units}

\newacronym{ctc}{CTC}{Common Test Conditions}

\newacronym{cnn}{CNN}{Convolutional Neural Network}
\newacronym{ccitt}{CCITT}{International Telegraph and Telephone Consultative Committee}

\newacronym{dvfs}{D\textsc{vfs}}{Dynamic Voltage and Frequency Scaling}
\newacronym{fd-soi}{FD-SOI}{Fully Depleted Silicon on Insulator}
\newacronym{dpm}{DPM}{Dynamic Power Management}

\newacronym{dst}{DST}{Discrete Sine Transform}
\newacronym{dct}{DCT}{Discrete Cosine Transform}

\newacronym{dbf}{DBF}{Deblocking Filter}
\newacronym{dea}{DEA}{Dominant Edge Assent}

\newacronym{dse}{DSE}{Design Space Exploration}

\newacronym{dpcm}{DPCM}{Differential Pulse-Code Modulation}
\newacronym{det}{$\Delta$ET}{$\Delta$ Encoding Time}

\newacronym{dt}{DT}{Decision Tree}

\newacronym{ec}{EC}{Entropy Coding}
\newacronym{er}{ER}{Energy Reduction}

\newacronym{eo}{EO}{Edges Offset}

\newacronym{frdcost}{$L_{FRD}$}{full \gls{rd} cost}
\newacronym{fps}{fps}{Frames per Second}
\newacronym{fvc}{FVC}{Future Video Coding}

\newacronym{fme}{FME}{Fractional Motion Estimation}

\newacronym{fdcs}{FDCS}{Fixed Depth Complexity Scaling}
\newacronym{ffmpeg}{FFmpeg}{Fast Forward \gls{mpeg}}
\newacronym{fifo}{FIFO}{First In First Out}

\newacronym{gof}{GOF}{Group of Frames}
\newacronym{gop}{GOP}{Group of Pictures}
\newacronym{gpu}{GPU}{Graphics Processing Unit}

\newacronym{hevc}{{HEVC}}{High Efficiency Video Coding}

\newacronym{hm}{HM}{\textsc{Hevc} test Model}

\newacronym{hd}{HD}{High Definition}

\newacronym{vceg}{VCEG}{IUT-T Video Coding Experts Group}

\newacronym{ittcc}{ITTCC}{International Telephone and Telegraph Consultative Committee}

\newacronym{ipe}{IPE}{Intra Picture Estimation}

\newacronym{ipp}{IPP}{Intra Picture Prediction}

\newacronym{ip}{IP}{Intra prediction}

\newacronym{iot}{IoT}{Internet of Things}

\newacronym{im}{IM}{Intra-mode prediction}
\newacronym{isp}{ISP}{Intra Sub-Partitions}

\newacronym{itu}{ITU}{International Telecommunication Union}
\newacronym{iso}{ISO}{International Organization for Standardization}
\newacronym{iec}{IEC}{International Electrotechnical Commission}
\newacronym{itut}{ITU-T}{International Telecommunication Union Telecommunication Standardization Sector}

\newacronym{jct-vc}{JCT-VC}{Joint Collaborative Team on Video Coding}
\newacronym{jvt}{JVT}{Joint Video Team}
\newacronym{jvet}{JVET}{Joint Video Experts Team}
\newacronym{jem}{JEM}{Joint Exploration Model}

\newacronym{kld}{KLD}{Kullback-Leibler Divergence}

\newacronym{ld}{LD}{Low Delay}
\newacronym{lb}{LB}{Low Delay B}
\newacronym{lp}{LP}{Low Delay P}
\newacronym{lcrdcost}{$J_{LRD}$}{low-complexity \gls{rd} cost}
\newacronym{lut}{LUT}{Look-Up Table}

\newacronym{lcb}{LCB}{Largest coding block}
\newacronym{lf}{$F_L$}{Learning Frame}
\newacronym{lgbm}{LGBM}{LightGBM}
\newacronym{lstm}{LSTM}{Long-Short Term Memory}

\newacronym{lmcs}{LMCS}{Luma Mapping with Chroma Scaling}
\newacronym{lfnst}{LFNST}{Low-Frequency Non-Separable Transform}

\newacronym{mpeg}{MPEG}{ISO/IEC Moving Picture Experts Group}
\newacronym{mse}{MSE}{Mean-Squared Error}
\newacronym{mep}{M\textsc{ep}}{Minimal Energy Point}
\newacronym{mep-ctu}{M\textsc{ep}-CTU}{Minimal Energy Point of \emph{\gls{ctu} level}}
\newacronym{mep-ip}{M\textsc{ep}-IP}{Minimal Energy Point of \emph{\gls{ip} level}}
\newacronym{msse}{M\textsc{sse}}{Minimization based on Search Space Exploration}
\newacronym{mcp}{M\textsc{cp}}{Minimal Cost Point}

\newacronym{mpm}{MPM}{Most Probable Mode}
\newacronym{mi}{MI}{Mutual Information}
\newacronym{msssr}{M\textsc{sssr}}{Multiple Smart Search Space Reduction}

\newacronym{ml}{ML}{Machine Learning}
\newacronym{mv}{MV}{Motion Vector}

\newacronym{mrf}{MRF}{Markov Random Field}

\newacronym{me}{ME}{Motion Estimation}
\newacronym{mc}{MC}{Motion Compensation}
\newacronym{mcc}{MCC}{Multi-Class Classifier}

\newacronym{mctdl}{MCTDL}{Motion-Compensated Tree Depth Limitation}

\newacronym{mts}{MTS}{Multiple Transform Selection}
\newacronym{mtt}{MTT}{Multi-Type Tree}
\newacronym{mrl}{MRL}{Multiple Reference Line}
\newacronym{mos}{MOS}{Mean Opinion Score}
\newacronym{movie}{MOVIE}{MOtion-tuned Video Integrity Evaluation}

\newacronym{msssim}{MS-SSIM}{Multi-Scale Structural Similarity}

\newacronym{mvd}{MVD}{Motion Vector Differences}
\newacronym{mvp}{MVP}{Motion Vector Prediction}

\newacronym{mip}{MIP}{Matrix-Based Intra Prediction}

\newacronym{ni}{NI}{National Instrument}

\newacronym{ntsc}{NTSC}{National Television System Committee}

\newacronym{nmos}{NMOS}{N-type Metal Oxide Semiconductor}
\newacronym{net}{NET}{Normalized Encoding Time}
\newacronym{niqe}{NIQE}{Natural Image Quality Evaluator}
\newacronym{nal}{NAL}{Network Abstraction Layer}

\newacronym{os}{OS}{Operating System}

\newacronym{pu}{PU}{Prediction Unit}
\newacronym{pb}{PB}{Prediction Block}

\newacronym{papi}{PAPI}{Performance Application Programming Interface}

\newacronym{psnr}{P\textsc{snr}}{Peak Signal-to-Noise Ratio}
\newacronym{pmos}{PMOS}{P-type Metal Oxide Semiconductor}

\newacronym{pdf}{PDF}{Probability Density Function}

\newacronym{pcci}{PCCI}{Percentage of Correctly Classify Instances}

\newacronym{pid}{PID}{Proportional–Integral–Derivative}

\newacronym{pi}{PI}{Proportional–Integral}

\newacronym{pal}{PAL}{Phase Alternating Line}
\newacronym{poc}{POC}{Picture Order Count}
\newacronym{q}{Q}{Quantization}
\newacronym{q-1}{Q\textsuperscript{-1}}{Inverse Quantization}
\newacronym{qp}{QP}{Quantization Parameter}

\newacronym{qoe}{QoE}{Quality of Experience}

\newacronym{qt}{QT}{Quad-Tree}

\newacronym{qtbt}{QTBT}{Quad-Tree plus Binary-Tree}
\newacronym{qtbttt}{QTBTTT}{Quad-Tree plus Binary-Tree plus Ternary-Tree}

\newacronym{rdo}{RDO}{Rate-Distortion Optimization}
\newacronym{rapl}{R\textsc{apl}}{Running Average Power Limit}

\newacronym{rap}{RAP}{Random Access Point}

\newacronym{rdoq}{RDOQ}{Rate-Distortion Optimization Quantization}

\newacronym{rd}{RD}{Rate-Distortion}

\newacronym{rqt}{RQT}{Residual Quad-Tree}

\newacronym{ra}{RA}{Random Access}

\newacronym{rmd}{RMD}{Rough Mode Decision}

\newacronym{rsq}{$Rsq$}{Coefficient of Determination}

\newacronym{rcdm}{\textsc{Rcdm}}{Refined \gls{ctu} Depth Map}

\newacronym{rc}{RC}{Rate-Control}

\newacronym{roc}{ROC}{Receiver Operating Characteristic}

\newacronym{rpr}{RPR}{Reference Picture Resampling}

\newacronym{sad}{SAD}{Sum of Absolute Differences}

\newacronym{sao}{SAO}{Sample Adaptive Offset}
\newacronym{simd}{S\textsc{imd}}{Single Instruction Multiple Data}
\newacronym{sssr}{S\textsc{ssr}}{Smart Search Space Reduction}
\newacronym{snr}{SNR}{Signal-to-Noise Ratio}

\newacronym{satd}{SATD}{Sum of Absolute Transformed Differences}
\newacronym{si}{SI}{Spatial Information}

\newacronym{svm}{SVM}{Support Vector Machine}

\newacronym{scb}{SCB}{Smallest coding block}

\newacronym{smp}{SMP}{Symmetric Motion Partition}
\newacronym{amp}{AMP}{Asymmetric Motion Partition}

\newacronym{sr}{SR}{Search Range}

\newacronym{scc}{SCC}{Subjective-driven Complexity Control}

\newacronym{sse}{SSE}{Sum of Squared Errors}
\newacronym{ssd}{SSD}{Sum of Squared Differences}
\newacronym{ssim}{SSIM}{Structural SIMilarity}
\newacronym{srr}{SRR}{SSIM Reduced-Reference}
\newacronym{sd}{SD}{Standard Definition}

\newacronym{tb}{TB}{Transform Block}
\newacronym{tu}{TU}{Transform Unit}
\newacronym{t}{T}{Transform}
\newacronym{t-1}{T\textsuperscript{-1}}{Inverse Transform}

\newacronym{ti}{TI}{Temporal Information}

\newacronym{td}{TD}{Top-Down}
\newacronym{tr}{TR}{Time Reduction}

\newacronym{tstd}{Tstd}{Time Standard Deviation}

\newacronym{tt}{TT}{Ternary-Tree}
\newacronym{uhd}{UHD}{Ultra High Definition}
\newacronym{vdcs}{VDCS}{Variable Depth Complexity Scaling}

\newacronym{vvc}{VVC}{Versatile Video Coding}

\newacronym{vtm}{VTM}{\gls{vvc} Test Model}

\newacronym{vr}{VR}{Virtual Reality}

\newacronym{vmaf}{VMAF}{Video Multimethod Assessment Fusion}

\newacronym{weka}{WEKA}{Waikato Environment for Knowledge Analysis}





%% file: tip_2020_tissier.bbl
\begin{thebibliography}{10}
\providecommand{\url}[1]{#1}
\csname url@samestyle\endcsname
\providecommand{\newblock}{\relax}
\providecommand{\bibinfo}[2]{#2}
\providecommand{\BIBentrySTDinterwordspacing}{\spaceskip=0pt\relax}
\providecommand{\BIBentryALTinterwordstretchfactor}{4}
\providecommand{\BIBentryALTinterwordspacing}{\spaceskip=\fontdimen2\font plus
\BIBentryALTinterwordstretchfactor\fontdimen3\font minus
  \fontdimen4\font\relax}
\providecommand{\BIBforeignlanguage}[2]{{%
\expandafter\ifx\csname l@#1\endcsname\relax
\typeout{** WARNING: IEEEtran.bst: No hyphenation pattern has been}%
\typeout{** loaded for the language `#1'. Using the pattern for}%
\typeout{** the default language instead.}%
\else
\language=\csname l@#1\endcsname
\fi
#2}}
\providecommand{\BIBdecl}{\relax}
\BIBdecl

\bibitem{cisco_cisco_2019}
Cisco, ``Cisco visual networking index: Forecast and trends, 2017-2022,'' in
  \emph{\url{https://www.cisco.com/c/en/us/solutions/collateral/service-provider/
  visual-networking-index-vni/ white-paper-c11-741490.html}}, Feb. 2019.

\bibitem{9503377}
B.~Bross, Y.~Wang, Y.~Ye, S.~Liu, J.~Chen, G.~Sullivan, and J.~Ohm, ``Overview
  of the versatile video coding (vvc) standard and its applications,''
  \emph{IEEE Transactions on Circuits and Systems for Video Technology},
  vol.~31, no.~10, pp. 3736--3764, 2021.

\bibitem{9689950}
W.~Hamidouche, T.~Biatek, M.~Abdoli, E.~François, F.~Pescador,
  M.~Radosavljević, D.~Menard, and M.~Raulet, ``Versatile video coding
  standard: A review from coding tools to consumers deployment,'' \emph{IEEE
  Consumer Electronics Magazine}, vol.~11, no.~5, pp. 10--24, 2022.

\bibitem{garcia-lucas_rate-distortioncomplexity_2020}
D.~García-Lucas, G.~Cebrián-Márquez, and P.~Cuenca,
  ``Rate-distortion/complexity analysis of {HEVC}, {VVC} and {AV}1 video
  codecs,'' \emph{Multimedia Tools and Applications}.

\bibitem{bossen_ahg_2020}
F.~Bossen, X.~Li, and K.~Suehring, ``{AHG} report: Test model software
  development ({AHG}3),'' \emph{{JVET}-T0003}.

\bibitem{huang_block_2021}
Y.~Huang, J.~An, H.~Huang, X.~Li, S.~Hsiang, K.~Zhang, H.~Gao, J.~Ma, and
  O.~Chubach, ``Block partitioning structure in the vvc standard,'' \emph{IEEE
  Transactions on Circuits and Systems for Video Technology}, vol.~31, no.~10,
  pp. 3818--3833, 2021.

\bibitem{francois_vvc_nodate}
E.~François, M.~Kerdranvat, R.~Jullian, and C.~Chevance, ``{VVC} {PER}-{TOOL}
  {PERFORMANCE} {EVALUATION} {COMPARED} {TO} {HEVC},'' p.~14.

\bibitem{saldanha_complexity_2020}
M.~Saldanha, G.~Sanchez, C.~Marcon, and L.~Agostini, ``Complexity analysis of
  {VVC} intra coding,'' in \emph{2020 {IEEE} International Conference on Image
  Processing ({ICIP})}.\hskip 1em plus 0.5em minus 0.4em\relax {IEEE}, pp.
  3119--3123.

\bibitem{tissier_complexity_2019}
A.~Tissier, A.~Mercat, T.~Amestoy, W.~Hamidouche, J.~Vanne, and D.~Menard,
  ``Complexity reduction opportunities in the future {VVC} intra encoder,'' in
  \emph{2019 {IEEE} 21st International Workshop on Multimedia Signal Processing
  ({MMSP})}.\hskip 1em plus 0.5em minus 0.4em\relax {IEEE}, pp. 1--6.

\bibitem{correa_fast_2015}
G.~Correa, P.~Assuncao, L.~Agostini, and L.~da~Silva~Cruz, ``Fast hevc encoding
  decisions using data mining,'' \emph{IEEE Transactions on Circuits and
  Systems for Video Technology}, vol.~25, no.~4, pp. 660--673, 2015.

\bibitem{li_accelerate_2020}
T.~Li, M.~Xu, X.~Deng, and L.~Shen, ``Accelerate {CTU} partition to real time
  for {HEVC} encoding with complexity control,'' \emph{{IEEE} Transactions on
  Image Processing}, vol.~29, pp. 7482--7496.

\bibitem{mercat_machine_2018}
A.~Mercat, F.~Arrestier, M.~Pelcat, W.~Hamidouche, and D.~Menard, ``Machine
  learning based choice of characteristics for the one-shot determination of
  the {HEVC} intra coding tree,'' in \emph{2018 Picture Coding Symposium
  ({PCS})}, pp. 263--267.

\bibitem{6862903}
L.~Shen, Z.~Zhang, and Z.~Liu, ``Effective cu size decision for hevc
  intracoding,'' \emph{IEEE Transactions on Image Processing}, vol.~23, no.~10,
  pp. 4232--4241, 2014.

\bibitem{6778776}
------, ``Adaptive inter-mode decision for hevc jointly utilizing inter-level
  and spatiotemporal correlations,'' \emph{IEEE Transactions on Circuits and
  Systems for Video Technology}, vol.~24, no.~10, pp. 1709--1722, 2014.

\bibitem{paul_speeding_2020}
S.~Paul, A.~Norkin, and A.~Bovik, ``Speeding up {VP}9 intra encoder with
  hierarchical deep learning based partition prediction,'' \emph{{IEEE}
  Transactions on Image Processing}, vol.~29, pp. 8134--8148.

\bibitem{xu_reducing_2018}
M.~Xu, T.~Li, Z.~Wang, X.~Deng, R.~Yang, and Z.~Guan, ``Reducing complexity of
  {HEVC}: A deep learning approach,'' \emph{{IEEE} Transactions on Image
  Processing}, vol.~27, no.~10, pp. 5044--5059.

\bibitem{wieckowski_fast_2019}
A.~Wieckowski, J.~Ma, H.~Schwarz, D.~Marpe, and T.~Wiegand, ``Fast partitioning
  decision strategies for the upcoming versatile video coding ({VVC})
  standard,'' in \emph{2019 {IEEE} International Conference on Image Processing
  ({ICIP})}, pp. 4130--4134.

\bibitem{wang_probabilistic_2018}
Z.~Wang, S.~Wang, J.~Zhang, S.~Wang, and S.~Ma, ``Probabilistic decision based
  block partitioning for future video coding,'' \emph{{IEEE} Transactions on
  Image Processing}, vol.~27, no.~3, pp. 1475--1486.

\bibitem{amestoy_tunable_2020}
T.~Amestoy, A.~Mercat, W.~Hamidouche, D.~Menard, and C.~Bergeron, ``Tunable
  {VVC} frame partitioning based on lightweight machine learning,''
  \emph{{IEEE} Transactions on Image Processing}, vol.~29, pp. 1313--1328.

\bibitem{9690615}
K.~Choi, T.~V. Le, Y.~Choi, and J.~Y. Lee, ``Low-complexity intra coding in
  versatile video coding,'' \emph{IEEE Transactions on Consumer Electronics},
  vol.~68, no.~2, pp. 119--126, 2022.

\bibitem{9328173}
X.~Dong, L.~Shen, M.~Yu, and H.~Yang, ``Fast intra mode decision algorithm for
  versatile video coding,'' \emph{IEEE Transactions on Multimedia}, vol.~24,
  pp. 400--414, 2022.

\bibitem{li_deepqtmt_2020}
T.~Li, M.~Xu, R.~Tang, Y.~Chen, and Q.~Xing, ``Deepqtmt: A deep learning
  approach for fast qtmt-based cu partition of intra-mode vvc,'' \emph{IEEE
  Transactions on Image Processing}, vol.~30, pp. 5377--5390, 2021.

\bibitem{tissier_cnn_2020}
A.~Tissier, W.~Hamidouche, J.~Vanne, F.~Galpin, and D.~Menard, ``{CNN} oriented
  complexity reduction of {VVC} intra encoder,'' in \emph{2020 {IEEE}
  International Conference on Image Processing ({ICIP})}.\hskip 1em plus 0.5em
  minus 0.4em\relax {IEEE}, pp. 3139--3143.

\bibitem{wang_extended_2020}
M.~Wang, J.~Li, L.~Zhang, K.~Zhang, H.~Liu, S.~Wang, S.~Kwong, and S.~Ma,
  ``Extended coding unit partitioning for future video coding,'' \emph{{IEEE}
  Transactions on Image Processing}, vol.~29, pp. 2931--2946.

\bibitem{zhang_fast_2020}
Q.~Zhang, Y.~Wang, L.~Huang, and B.~Jiang, ``Fast {CU} partition and intra mode
  decision method for h.266/{VVC},'' \emph{{IEEE} Access}, vol.~8, pp.
  117\,539--117\,550.

\bibitem{fu_two-stage_2019}
T.~Fu, H.~Zhang, F.~Mu, and H.~Chen, ``Two-stage fast multiple transform
  selection algorithm for {VVC} intra coding,'' in \emph{2019 {IEEE}
  International Conference on Multimedia and Expo ({ICME})}.\hskip 1em plus
  0.5em minus 0.4em\relax {IEEE}, pp. 61--66.

\bibitem{biao_min_fast_2015}
{Biao Min} and R.~Cheung, ``A fast {CU} size decision algorithm for the {HEVC}
  intra encoder,'' \emph{{IEEE} Transactions on Circuits and Systems for Video
  Technology}, vol.~25, no.~5, pp. 892--896.

\bibitem{wang_effective_2017}
Z.~Wang, S.~Wang, J.~Zhang, S.~Wang, and S.~Ma, ``Effective quadtree plus
  binary tree block partition decision for future video coding,'' in \emph{2017
  Data Compression Conference ({DCC})}, pp. 23--32.

\bibitem{jin_cnn_2017}
Z.~Jin, P.~An, L.~Shen, and C.~Yang, ``{CNN} oriented fast {QTBT} partition
  algorithm for {JVET} intra coding,'' in \emph{2017 {IEEE} Visual
  Communications and Image Processing ({VCIP})}, pp. 1--4.

\bibitem{yang_low_2019}
H.~Yang, L.~Shen, X.~Dong, Q.~Ding, P.~An, and G.~Jiang, ``Low-complexity ctu
  partition structure decision and fast intra mode decision for versatile video
  coding,'' \emph{IEEE Transactions on Circuits and Systems for Video
  Technology}, vol.~30, no.~6, pp. 1668--1682, 2020.

\bibitem{chen_fast_2020}
F.~Chen, Y.~Ren, Z.~Peng, G.~Jiang, and X.~Cui, ``A fast {CU} size decision
  algorithm for {VVC} intra prediction based on support vector machine,''
  \emph{Multimedia Tools and Applications}.

\bibitem{zhao_adaptive_2020}
J.~Zhao, Y.~Wang, and Q.~Zhang, ``Adaptive {CU} split decision based on deep
  learning and multifeature fusion for h.266/{VVC},'' \emph{Scientific
  Programming}, vol. 2020, pp. 1--11.

\bibitem{saldanha_configurable_2021}
M.~Saldanha, G.~Sanchez, C.~Marcon, and L.~Agostini, ``Configurable fast block
  partitioning for vvc intra coding using light gradient boosting machine,''
  \emph{IEEE Transactions on Circuits and Systems for Video Technology}, pp.
  1--1, 2021.

\bibitem{lei_look-ahead_2019}
M.~Lei, F.~Luo, X.~Zhang, S.~Wang, and S.~Ma, ``Look-ahead prediction based
  coding unit size pruning for {VVC} intra coding,'' in \emph{2019 {IEEE}
  International Conference on Image Processing ({ICIP})}, pp. 4120--4124.

\bibitem{bross_multiple_2018}
B.~Bross, P.~Keydel, H.~Schwarz, D.~Marpe, T.~Wiegand, L.~Zhao, X.~Zhao, X.~Li,
  S.~Liu, Y.~Chang, H.~Jiang, P.~Lin, C.~Kuo, C.~Lin, and C.~Lin, ``Multiple
  reference line intra prediction,'' \emph{{JVET}-L0283}.

\bibitem{he_deep_2015}
\BIBentryALTinterwordspacing
K.~He, X.~Zhang, S.~Ren, and J.~Sun, ``Deep residual learning for image
  recognition,'' \emph{{arXiv}:1512.03385 [cs]}. [Online]. Available:
  \url{http://arxiv.org/abs/1512.03385}
\BIBentrySTDinterwordspacing

\bibitem{tan_efficientnet_2019}
\BIBentryALTinterwordspacing
M.~Tan and Q.~V. Le, ``{EfficientNet} rethinking model scaling for
  convolutional neural networks,'' \emph{{arXiv}:1905.11946 [cs, stat]}.
  [Online]. Available: \url{http://arxiv.org/abs/1905.11946}
\BIBentrySTDinterwordspacing

\bibitem{simonyan_very_2015}
\BIBentryALTinterwordspacing
K.~Simonyan and A.~Zisserman, ``Very deep convolutional networks for
  large-scale image recognition,'' \emph{{arXiv}:1409.1556 [cs]}. [Online].
  Available: \url{http://arxiv.org/abs/1409.1556}
\BIBentrySTDinterwordspacing

\bibitem{ke_lightgbm_2017}
G.~Ke, Q.~Meng, T.~Finley, T.~Wang, W.~Chen, W.~Ma, Q.~Ye, and T.~Liu,
  ``{LightGBM}: A highly efficient gradient boosting decision tree,''
  \emph{Advances in Neural Information Processing Systems 30}, p.~9.

\bibitem{agustsson_ntire_2017}
E.~Agustsson and R.~Timofte, ``{NTIRE} 2017 challenge on single image
  super-resolution: Dataset and study,'' in \emph{2017 {IEEE} Conference on
  Computer Vision and Pattern Recognition Workshops ({CVPRW})}.\hskip 1em plus
  0.5em minus 0.4em\relax {IEEE}, pp. 1122--1131.

\bibitem{makov_dataset_2019}
\BIBentryALTinterwordspacing
E.~Makov, \emph{Dataset image 4k}. [Online]. Available:
  \url{https://www.kaggle.com/evgeniumakov/images4k}
\BIBentrySTDinterwordspacing

\bibitem{jtc_call_2020}
I.~Jtc and I.-T. Sg, ``Call for evidence on learning-based image coding
  technologies ({JPEG} {AI}),'' p.~15.

\bibitem{hasinoff_burst_2016}
S.~Hasinoff, D.~Sharlet, R.~Geiss, A.~Adams, J.~Barron, F.~Kainz, J.~Chen, and
  M.~Levoy, ``Burst photography for high dynamic range and low-light imaging on
  mobile cameras,'' \emph{{ACM} Transactions on Graphics}, vol.~35, no.~6, pp.
  1--12.

\bibitem{timofte_ntire_nodate}
R.~e.~a. Timofte, ``{NTIRE} 2017 challenge on single image super-resolution:
  Methods and results,'' p.~12.

\bibitem{chollet_et_al_keras_2015}
\BIBentryALTinterwordspacing
F.~Chollet~et al., \emph{Keras}. [Online]. Available: \url{https://keras.io}
\BIBentrySTDinterwordspacing

\bibitem{abadi_et_al_tensorflow_nodate}
F.~Abadi~et al., ``{TensorFlow}: Large-scale machine learning on heterogeneous
  distributed systems,'' p.~19.

\bibitem{kingma_adam_2017}
\BIBentryALTinterwordspacing
D.~Kingma and J.~Ba, ``Adam: A method for stochastic optimization,''
  \emph{{arXiv}:1412.6980 [cs]}. [Online]. Available:
  \url{http://arxiv.org/abs/1412.6980}
\BIBentrySTDinterwordspacing

\bibitem{bossen_jvet_2019}
F.~Bossen, J.~Boyce, K.~Suehring, X.~Li, and V.~Seregin, ``{JVET} common test
  conditions and software reference configurations for {SDR} video,''
  \emph{{JVET} document, {JVET}-M1010}.

\bibitem{bjontegaard_calculation_2001}
G.~Bjontegaard, ``Calculation of average {PSNR} differences between
  {RD}-curves,'' \emph{{VCEG}-M33}.

\bibitem{hermann_frugally_2018}
\BIBentryALTinterwordspacing
T.~Hermann, ``Frugally deep.'' [Online]. Available:
  \url{https://github.com/Dobiasd/frugally-deep}
\BIBentrySTDinterwordspacing

\end{thebibliography}
